\RequirePackage[2020-02-01]{latexrelease}
\documentclass[twocolumn]{article} 

\usepackage{preprint}

\usepackage{amsmath, amsthm, amssymb, amsfonts}

\usepackage[numbers,square]{natbib}

\usepackage[utf8]{inputenc}	
\usepackage[T1]{fontenc}	
\usepackage{xcolor}		
\usepackage{colortbl}
\usepackage[colorlinks = true,
            linkcolor = purple,
            urlcolor  = blue,
            citecolor = cyan,
            anchorcolor = black]{hyperref}	
\usepackage{subcaption}
\usepackage{graphicx}
\usepackage{soul}
\usepackage{enumitem}
\usepackage{booktabs} 		
\usepackage{nicefrac}		
\usepackage{microtype}		
\usepackage{lineno}		
\usepackage{float}			

\usepackage{lipsum}		

\usepackage{newfloat}
\DeclareFloatingEnvironment[name={Supplementary Figure}]{suppfigure}
\usepackage{sidecap}
\sidecaptionvpos{figure}{c}

\usepackage{titlesec}
\titlespacing\section{0pt}{12pt plus 3pt minus 3pt}{1pt plus 1pt minus 1pt}
\titlespacing\subsection{0pt}{10pt plus 3pt minus 3pt}{1pt plus 1pt minus 1pt}
\titlespacing\subsubsection{0pt}{8pt plus 3pt minus 3pt}{1pt plus 1pt minus 1pt}

\usepackage{tikz,xcolor,hyperref}
\makeatletter
 \DeclareRobustCommand\ref{%
    \@ifstar\@refstar\T@ref
  }%
  \DeclareRobustCommand\pageref{%
    \@ifstar\@pagerefstar\T@pageref
  }%
 \makeatother
\definecolor{lime}{HTML}{A6CE39}
\DeclareRobustCommand{\orcidicon}{
	\begin{tikzpicture}
	\draw[lime, fill=lime] (0,0)
	circle [radius=0.16]
	node[white] {{\fontfamily{qag}\selectfont \tiny ID}};
	\draw[white, fill=white] (-0.0625,0.095)
	circle [radius=0.007];
	\end{tikzpicture}
	\hspace{-2mm}
}
\foreach \x in {A, ..., Z}{\expandafter\xdef\csname orcid\x\endcsname{\noexpand\href{https://orcid.org/\csname orcidauthor\x\endcsname}
			{\noexpand\orcidicon}}
}

\title{CISCA and CytoDArk0: a Cell Instance Segmentation and Classification method for histo(patho)logical image Analyses and a new, open, Nissl-stained dataset for brain cytoarchitecture studies}%

\usepackage{xwatermark}
\newwatermark[firstpage,color=gray!90,angle=0,scale=0.28, xpos=0in,ypos=-5in]{*Corresponding Author: \texttt{vadoriv@lsbu.ac.uk}}

\usepackage{authblk}

\author[1]{Valentina Vadori}
\author[2]{Jean-Marie Graïc}
\author[2]{Antonella Peruffo}
\author[2]{Giulia Vadori}
\author[3]{Livio Finos}
\author[1]{Enrico Grisan}

\affil[1]{London South Bank University, School of Engineering, United Kingdom}
\affil[2]{University of Padova, Dept. of Comparative Biomedicine \& Food Science, Italy}
\affil[3]{University of Padova, Dept. of Statistical Sciences, Italy}

\begin{document}

\twocolumn[ 
  \begin{@twocolumnfalse} 

\maketitle

\begin{abstract}
Delineating and classifying individual cells in microscopy tissue images is inherently challenging yet remains essential for advancements in medical and neuroscientific research.  In this work, we propose a new deep learning framework, CISCA, for automatic cell instance segmentation and classification in histological slices. At the core of CISCA is a network architecture featuring a lightweight U-Net with three heads in the decoder. The first head classifies pixels into boundaries between neighboring cells, cell bodies, and background, while the second head regresses four distance maps along four directions. The outputs from the first and second heads are integrated through a tailored post-processing step, which ultimately produces the segmentation of individual cells. The third head enables the simultaneous classification of cells into relevant classes, if required. We demonstrate the effectiveness of our method using four datasets, including CoNIC, PanNuke, and MoNuSeg, which are publicly available H\&E-stained datasets that cover diverse tissue types and magnifications. In addition, we introduce CytoDArk0, the first annotated dataset of Nissl-stained histological images of the mammalian brain, containing nearly 40k annotated neurons and glia cells, aimed at facilitating advancements in digital neuropathology and brain cytoarchitecture studies. We evaluate CISCA against other state-of-the-art methods, demonstrating its versatility, robustness, and accuracy in segmenting and classifying cells across diverse tissue types, magnifications, and staining techniques. This makes CISCA well-suited for detailed analyses of cell morphology and efficient cell counting in both digital pathology workflows and brain cytoarchitecture research.
\end{abstract}
\vspace{0.5cm}

  \end{@twocolumnfalse} 
] 



\section{Introduction}
\label{sec:Intro}
``\textit{No matter how we twist and turn, we shall always come back to the cell}'' \citep{virchow1860cellular}. The German polymath Rudolf Virchow, often referred to as the \textit{father of modern pathology}, stated that all deviations from health in any disease invariably trace back to alterations within cells, that is, all \textit{pathology} ultimately is \textit{cellular pathology}. His groundbreaking insights laid the foundation for contemporary \textit{histopathology}, 
the diagnostic discipline that interprets cellular biology by examining tissue slides obtained through biopsies or surgical procedures under a microscope \citep{jahn2020digital}. This method is the established approach for diagnosing and grading various cancer types, including breast, prostate, colorectal, and skin cancers, and inflammatory diseases \citep{van2021deep}. 
In addition to providing diagnostic information, the scrutiny of multiple level features such as alterations in tissue architecture, extent of glandular formation, presence of mitosis or inflammatory cells, and nuclear atypia, allows to pinpoint deviations indicative of disease progression and aggressiveness for prognosis and treatment decisions \citep{srinidhi2021deep}. 

Over the past two decades, advancements in imaging technologies and storage devices have catalyzed the widespread digitization of histological glass slides, marking the beginning of an ongoing transition to \textit{digital pathology} \citep{aeffner2019introduction}. This transformation has offered numerous benefits, including improved accessibility to slides, enhanced collaboration among pathologists, and availability of remote consultation \citep{madabhushi2016image}. Digital pathology has also opened doors to computational image analysis, ushering in a new era of \textit{computational pathology} \citep{hosseini2024comput}. 
Recently, there has been a focused shift in research towards deep learning (DL) methodologies, distinguished by their capability to learn representations with multiple levels of abstraction directly from raw data, bypassing the need for manually crafted features developed with expert assistance. With the growing availability of computing power and annotated histopathology datasets, which DL methods eagerly leverage, there is a progressive potential to enhance clinician decision-making through augmented intelligence. Such augmentation stands to enhance the efficiency, objectivity, and consistency of clinical decisions, to facilitate the exploration of novel biomarkers derived from subtle tissue and cellular traits, and to accurately predict therapy responses \citep{van2021deep}. 

Extensive research has focused on harnessing DL for detecting, segmenting, and classifying \textit{histological primitives}, especially individual cells or sub-cellular compartments. These advancements are crucial for diagnostic and prognostic assessments of various diseases, which heavily depend on factors such as cell quantity, size, appearance, morphology, and spatial arrangement \citep{lee2021deep}. In particular, the process of \textit{cell instance segmentation}, which involves identifying single cells and outlining their boundaries, along with \textit{cell classification} to recognize cell types, plays an important role in extracting interpretable biomarkers that characterize tissue microenvironment and can be used in downstream clinical tasks, such as cancer grading \citep{shaban2020context}, origin identification for cancers \citep{lu2021ai}, improved patient stratification \citep{dong2014computational, yu2016predicting, wulczyn2020deep}.

In the realm of biomedical research extending beyond clinical practice, the field of \textit{comparative brain cytoarchitecture} could greatly benefit from a DL-based, automated and reproducible approach to segmentation and classification of individual cells within whole slide images (WSI). Once cells are segmented and classified, the detailed organization of brain cells—including their types, densities, and spatial arrangements—known as \textit{brain cytoarchitecture}, can be precisely quantified. The observations gleaned have the potential to enhance our understanding of the pathogenesis of various human neurodegenerative and neuroinflammatory disorders such as Parkinson's disease \citep{cave2016cyto}, schizophrenia \citep{bakhshi2015neuropathology}, autism spectrum disorder (ASD) \citep{varghese2017autism}, and systemic lupus erythematosus \citep{graic2023cytoarchitectureal}, and to advance veterinary pathology \citep{zuraw2022whole}. For example,  \citet{falcone2021neuronal} investigated layer-specific changes in cell types numbers in the prefrontal cortex using Nissl-stained tissue from postmortem human brains of individuals with ASD to determine alterations that can affect cortical circuit regulation and behavioral outcomes. However, quantification was manually performed on small subsets of regions of interest. Once an initial effort is dedicated to establishing an annotated dataset, DL techniques could facilitate meaningful comparisons between postmortem histology images from healthy and diseased subjects, mitigating the limitations of semi-quantitative assessments, which are prone to region selection bias and inter-rater variability, time-consuming, and difficult to scale \citep{pansuwan2023accurate}.

Large-scale comparisons of cytoarchitecture across diverse species, age groups, or sexes can also potentially provide researchers with valuable insights into the developmental trajectories and evolutionary adaptations shaping brain complexity and information-processing capacity \citep{AMUNTS20071061}. For example, \citet{schenker2008comparative} employed a semiautomated method to investigate the neural underpinnings of linguistic functions.
Their study compared the human Broca's region with its homologs in great apes using quantitative analysis of cytoarchitecture from Nissl-stained histological sections. The researchers discovered that the connectivity space in humans is notably greater, a finding reinforced by \citet{palomero2019differences} and subsequent studies examining fiber tracts and cortical minicolumn spacing. These variations in brain microarchitecture, alongside other anatomical differences, contribute to understanding why non-human primates lack language abilities while humans possess them. To quantify connectivity space, \citet{schenker2008comparative} utilized the gray level index (GLI), representing the area fraction occupied by cells. However, GLI, which is usually computed via  thresholding techniques, does not differentiate between individual cell instances, thus limiting precise quantification of neuron numbers, layer distribution, and variations in cell-level features. These nuanced measures could enhance similar studies, potentially uncovering new insights.
For instance, \citet{graic2024age} represents the first study to apply quantitative multivariate analysis to assess cytoarchitectural changes in the brains of cetaceans across their lifespan. The cell instance segmentation algorithm detailed in \citet{vadori2023mr} was utilized to delineate individual cells and extract cell-level features from Nissl-stained histological images of the primary auditory cortex of new-born, adult and old bottlenose dolphins. Subsequent comparison of neuronal cell size, regularity, and density across cortical layers revealed subtle age-related changes in cytoarchitecture, primarily observed in interneurons located in the upper cortical layers. While this finding is not conclusive, it provides compelling evidence for the significance of these layers within the current thalamo-cortical input model for cetaceans and supports the hypothesis of ongoing connection development throughout a dolphin's (and possibly other cetaceans') lifespan.

Despite the research efforts invested over the last decade, accurately detecting and classifying individual cells in images can be quite challenging, particularly when cells overlap, touch, or display varied morphology, intensity, or appearance across different anatomical regions, conditions, and staining techniques. This complexity is further exacerbated by additional artifacts that can significantly impact the results. DL models have risen in prominence, outperforming traditional or shallow machine learning (ML) methods. Nonetheless, they require a significant amount of annotated data for effective training.
While open datasets exist for H\&E histological cells, to the best of our knowledge, there is no open dataset with single cell annotations for brain histological sections stained with the Nissl method, one of the most frequently used staining for cytoarchitecture studies  \citep{garcia2016distinction}.
With the work hereby presented, our aim is thus to make the following contributions:
\begin{enumerate}
\item We introduce CISCA, a novel and relatively lightweight processing framework designed for Cell Instance Segmentation and Classification, specifically aimed at the Analysis of whole slide histo(patho)logical images.
\item We evaluate CISCA on four distinct histological datasets, including the CoNIC challenge dataset of H\&E images of the colon and the PanNuke pan-cancer dataset, comparing its performance with respect to state-of-the-art (SOTA) methods and demonstrating its robustness and accuracy on different staining, magnifications and tissue types.
\item We release the first open annotated dataset (CytoDArk0) for cell instance segmentation in Nissl-stained 2D histological images of the brain. 
\end{enumerate}
CISCA and the CytoDArk0 dataset were created by our group to explore brain cytoarchitecture across various animal species, including cetaceans, primates, and ungulates. In the absence of an annotated dataset, we initially developed a simpler ML approach \citep{vadori2023mr} and iteratively annotated a first set of WSI patches. We transitioned to a DL paradigm with NCIS \citep{vadori2023ncis}, which served as the precursor to CISCA. Key adjustments were made to the previous method. We substituted the pre-trained EfficientNet-B5 \citep{tan2019efficientnet} backbone with a more lightweight and straightforward backbone to decrease the computational load while maintaining performance levels. The individual decoding branches were replaced with simpler heads that share most of the feature maps.
A new head was added to classify cells into different types. Additionally, the post-processing pipeline was enhanced to effectively manage both 40x and 20x magnifications. By sharing CISCA source code, model weights, and the CytoDArk0  dataset, we aim to foster collaboration among researchers and facilitate the creation of accurate tools for digital neuropathology applications and brain cytoarchitecture studies.

Our subsequent presentation is structured as follows: Section \ref{related} focuses on the recent literature for cell instance segmentation and classification; Section \ref{method} details CISCA methodology; Section \ref{datasets} describes the datasets; Section \ref{results} illustrates the results; Section \ref{conclusions} discusses the results and draws the conclusions.

\begin{figure*}%
\centering
\begin{subfigure}{.405\columnwidth}
\includegraphics[width=\columnwidth]{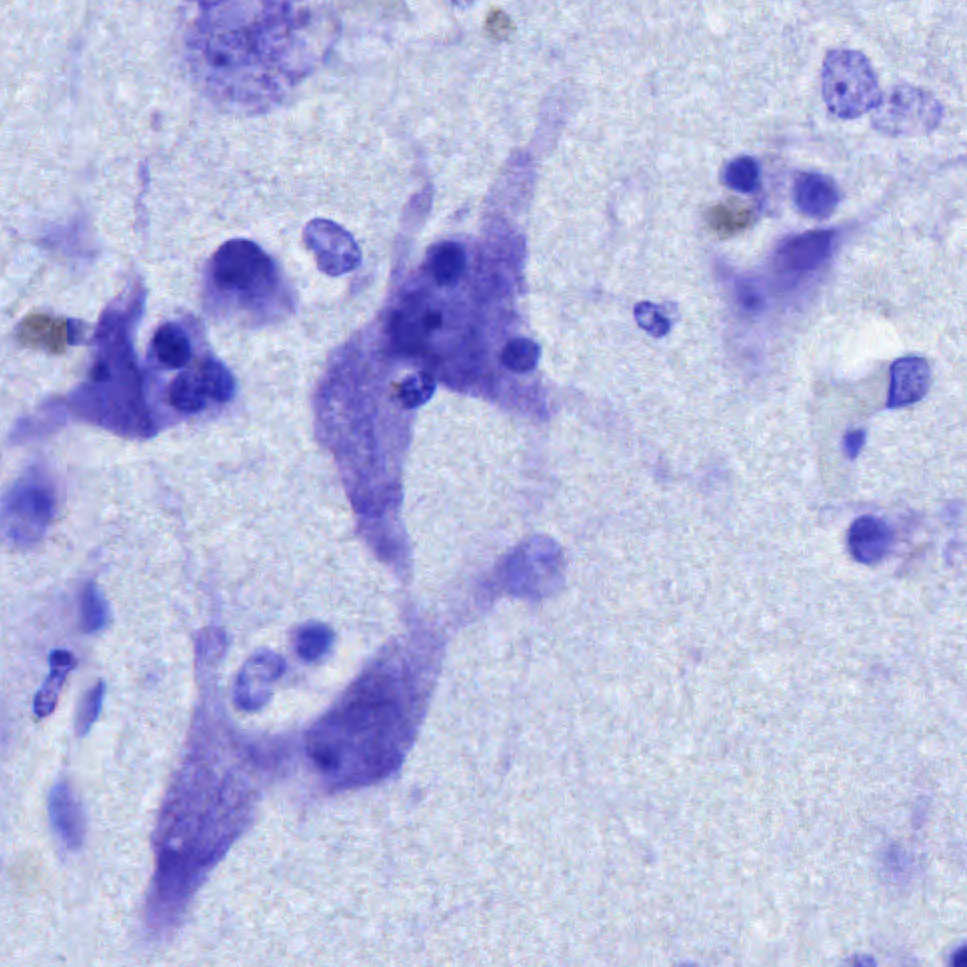}%
\caption{}%
\end{subfigure}\hfill%
\begin{subfigure}{.405\columnwidth}
\includegraphics[width=\columnwidth]{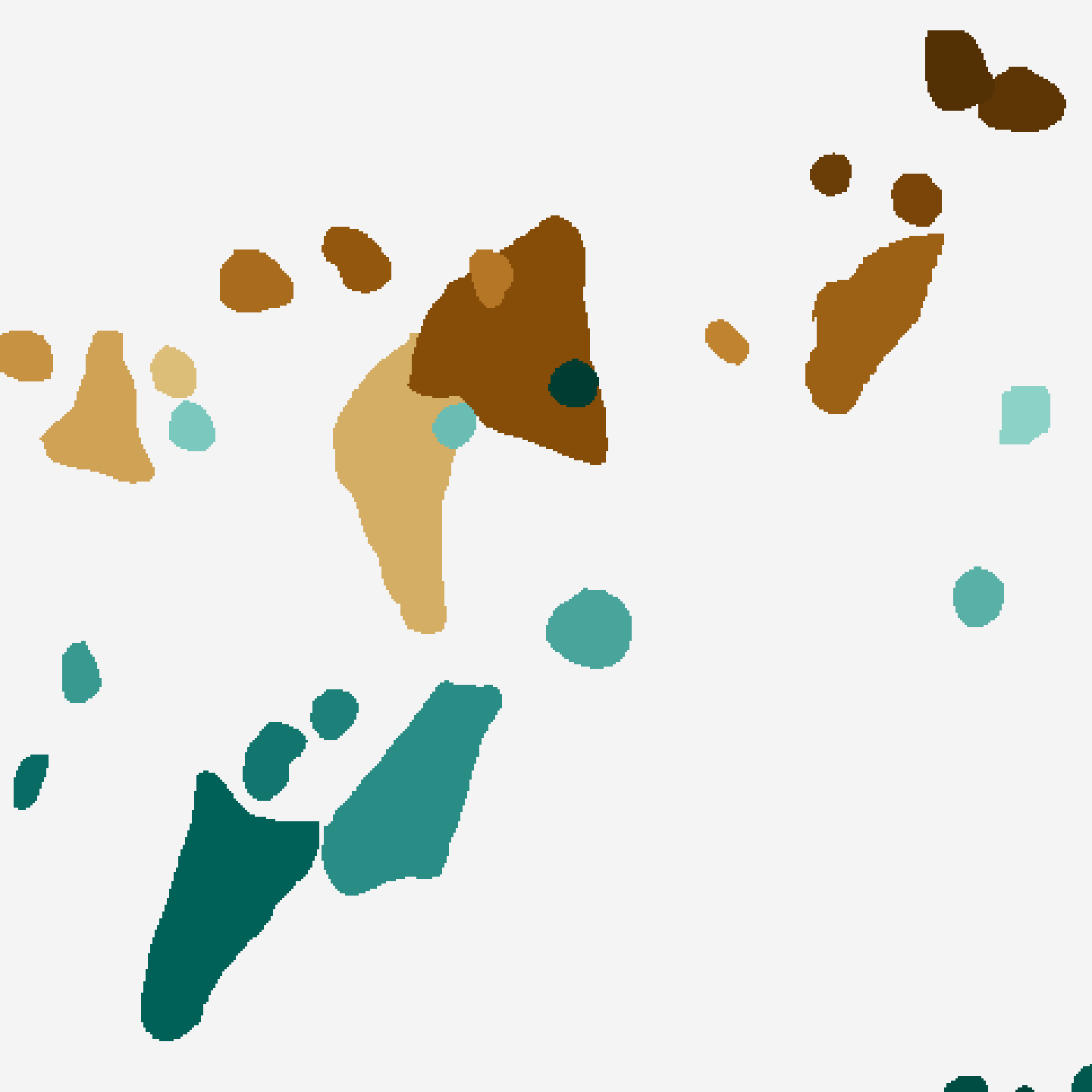}%
\caption{}%
\end{subfigure}\hfill%
\begin{subfigure}{.405\columnwidth}
\includegraphics[width=\columnwidth]{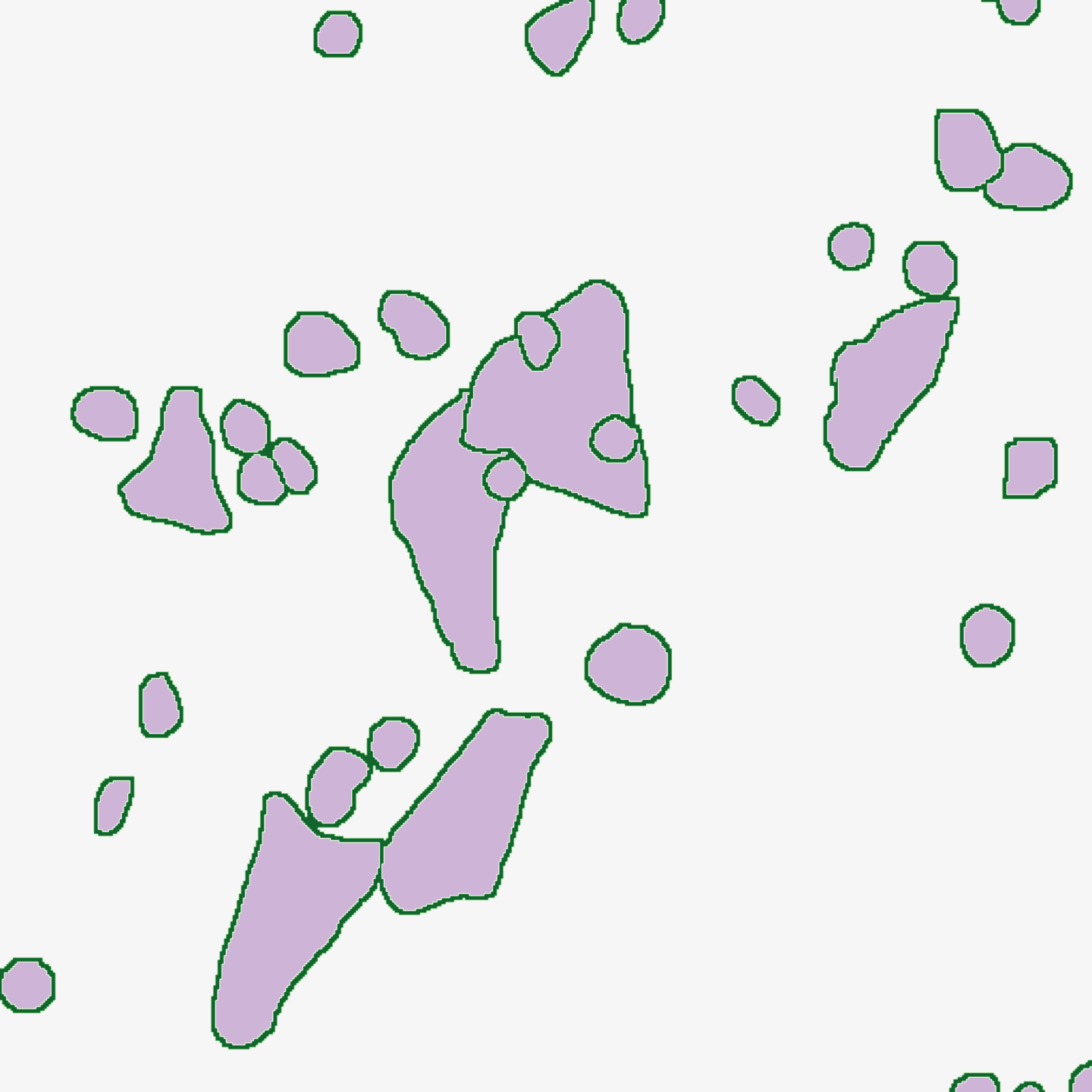}%
\caption{}%
\end{subfigure}\hfill%
\begin{subfigure}{.405\columnwidth}
\includegraphics[width=\columnwidth]{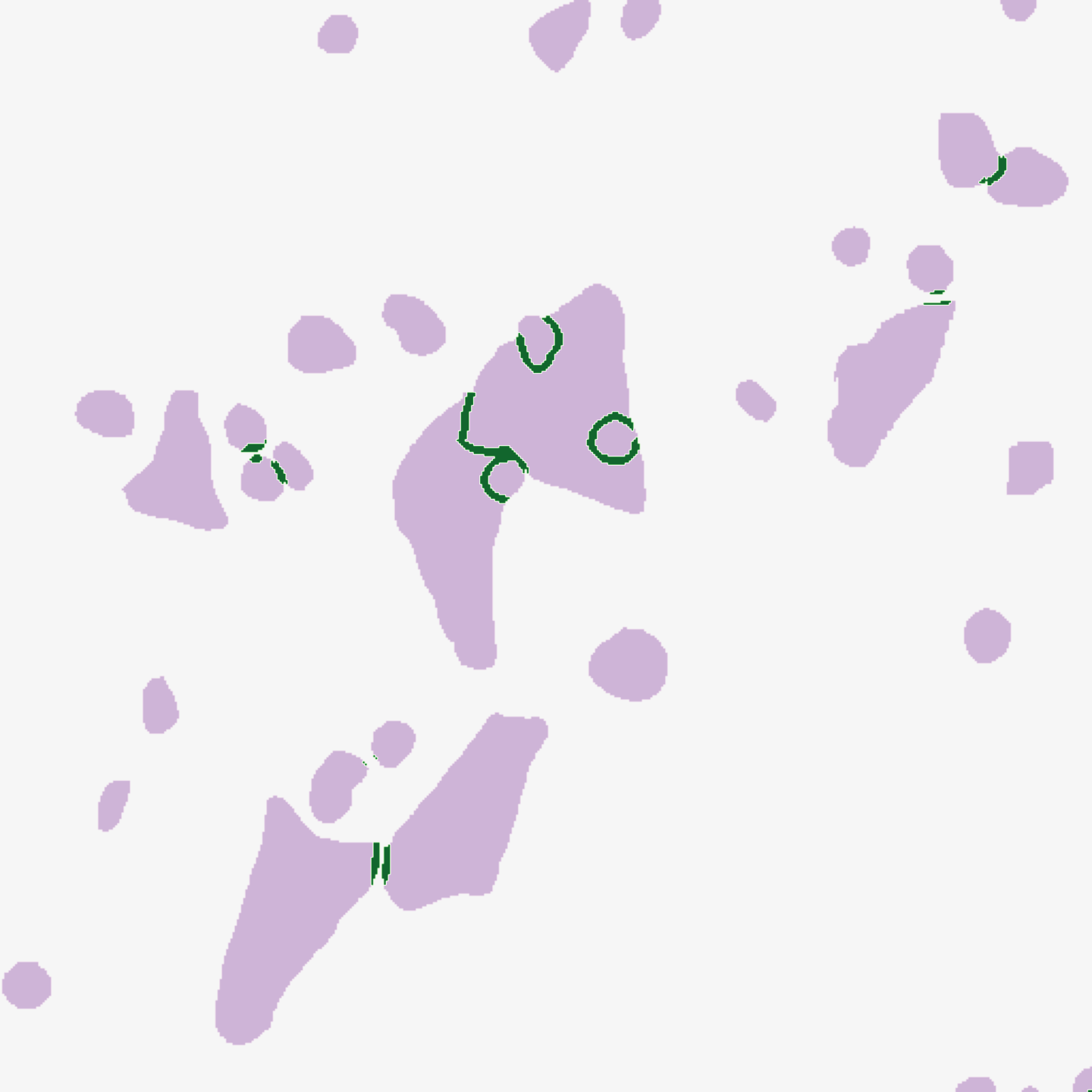}%
\caption{}%
\end{subfigure}\hfill%
\begin{subfigure}{.405\columnwidth}
\includegraphics[width=\columnwidth]{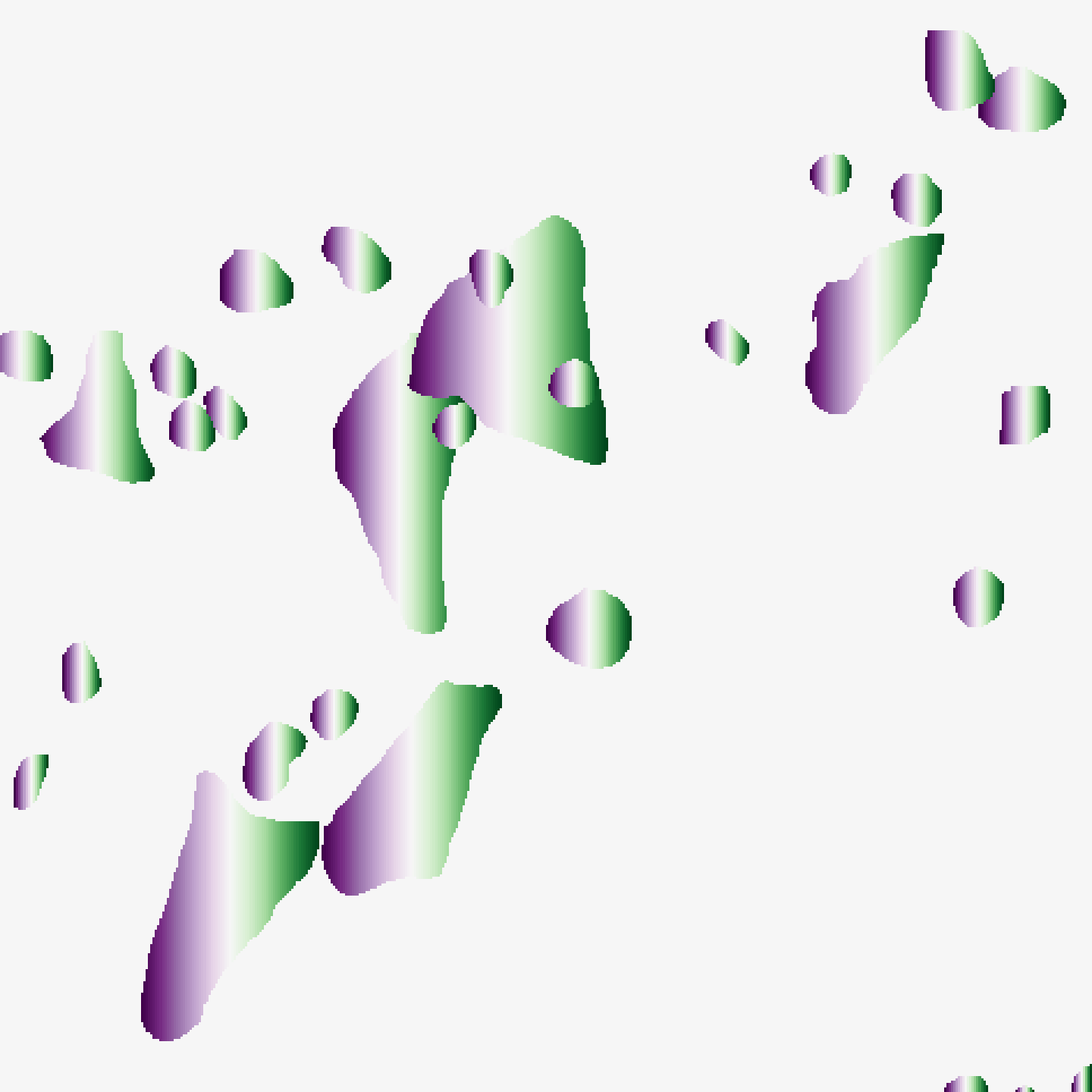}%
\caption{}%
\end{subfigure}
\caption{Illustration of relevant maps for the instance segmentation of cells a) Sample patch from the CytoDArk0  dataset. b) Instance segmentation label map (GT) c) Semantic segmentation map - $3$ pixel classes with all-round contours. d) Semantic segmentation map - $3$ pixel classes with the contours class limited to the boundary between cells that touch or are in close proximity. e) Example of a distance map along a specific direction as in Hover-Net \citep{graham2019hover} or in our proposed approach, CISCA.}
\label{40xPadova}
\end{figure*}

\section{Related Work}
\label{related}

Identifying and segmenting individual cells has long been a challenge, dating back to the early days of digitized microscopy images \citep{meijering2012cell}. Traditional approaches often relied on algorithms like intensity thresholding, morphological operations, watershed transforms, deformable models, clustering, and graph-based techniques \citep{XingRobust}. However, with the rise of DL, particularly convolutional neural networks (CNNs), significant advancements have been made in a wide range of machine vision tasks, including cell instance segmentation, where CNNs have outperformed traditional methods \citep{moen2019deep}. Despite international competitions highlighting various techniques \citep{caicedo2019nucleus, kumar2019multi, mavska2023cell}, a universally effective solution remains elusive. 

In recent years, the instance segmentation problem has predominantly been cast as a \textit{semantic segmentation} task, where the goal is to classify each pixel as either background (BG) or foreground. While this strategy is effective at separating isolated objects, it encounters difficulties with touching ones. Marker-controlled watershed can assist in separating individual instances, combined with techniques such as LoG filtering or radial symmetry-based voting to identify seed points \cite{kumar2019multi}. However, this approach struggles when cell shapes deviate from circular and boundaries between touching cells are unclear. To address this, a third pixel class, representing cell boundaries, has been introduced to better separate neighboring instances, as shown in Fig. \ref{40xPadova} c). Alternatively, \citet{wu2022general} propose a ternary classification approach that distinguishes BG, cell body, and inter-cell boundaries, similarly to Fig. \ref{40xPadova} d). \citet{pena2020j} exploit a fourth `gap' class for regions of the background near the borders of touching or closely positioned cells. To convert the semantic segmentation map into an instance segmentation map, a post-processing step is required, which assigns unique labels to each cell (see Fig. \ref{40xPadova} b)). This step typically uses morphological operations, conditional random fields or watershed algorithms. 

Most commonly, the semantic segmentation task is addressed via a U-Net-like CNN architecture \citep{ronneberger2015u}, which includes a contracting encoding path and an expansive decoding path. The encoding path captures context and reduces spatial resolution through convolutional and pooling layers, while the decoding path enables precise localization using upsampling layers and concatenating feature maps from the encoding path via \textit{skip connections}. Variations of this architecture often incorporate different convolutional encoders and innovative upsampling decoders. For instance, DCAN \citep{chen2017dcan} and CIA-Net \citep{zhou2019cia} use dual decoding branches to segment the cell body and boundaries separately, while \citet{cui2019deep} use a single branch to simultaneously predict BG, cell body, and boundaries. 

However, semantic segmentation alone often struggles with densely clustered cells or cells with diverse shapes, complicating their effective separation during post-processing. To address these challenges, \citet{naylor2018segmentation} propose reframing the problem as the \textit{regression} of a distance map, where each pixel value predicts its distance from the cell edges, naturally yielding a topographic landscape for the watershed algorithm to operate on. Building on this idea, several SOTA methods combine semantic segmentation with an auxiliary regression task. For example, Hover-Net \citep{graham2019hover} enhances the binary semantic segmentation task by regressing pixel distances from the cell centroid along the horizontal and vertical directions, as shown in Fig. \ref{40xPadova} e), using a U-Net-like architecture with an encoding path inspired by the pre-activated residual network \citep{he2016identity}. The distances are normalized and signed, ranging from $-1$ to $1$ within each cell. During the post-processing phase, Hover-Net exploits the differences in these distance values at cell boundaries (e.g., $-1$ vs. $1$) by applying Sobel operators to detect and separate the edges of touching cells. In a similar vein, CDNet \citep{he2021cdnet} regresses a discretized centripetal direction pointing towards the cell centroid for each pixel within a cell and exploits directional differences at cell boundaries to separate touching cells. 
CellPose \citep{stringer2021cellpose} employs a comparable centripetal direction method, but rather than using boundary differences, it tracks predicted directions  to guide pixels towards the cell centers and group them accordingly. 
Other methods, like DenseRes-Unet \citep{kiran2022denseres} and Mesmer \citep{greenwald2022whole}, regress an inner distance transform, capturing the distance of each pixel inside a cell to the cell centroid. 
Further, \citet{scherr2020cell} complement this with a map of the distance to the nearest neighbor.
Inspired by the UNETR architecture \citep{hatamizadeh2022unetr}, \citet{horst2024cellvit} propose CellViT, a U-Net-shaped architecture where the encoding path is replaced with a Vision Transformer (ViT). ViT splits the input image into a series of smaller patches, which are then flattened and linearly projected to create patch embeddings. These embeddings serve as input to the transformer encoder, which, differently from CNNs, consists of multiple blocks with alternating layers of multiheaded self-attention and multilayer perceptrons. Following Hover-Net \citep{graham2019hover}, CellViT utilizes two decoding paths: one for the binary semantic segmentation task and another for regressing normalized, signed distances from the cell centroids. To integrate tissue type information, they also introduce a tissue classification branch to guide the encoder's learning process.
StarDist \citep{schmidt2018cell} models cell shapes as star-convex polygons with a fixed number of radii. It uses a dense probability map to detect individual cells and regresses a distance map for each radius, where each pixel’s value represents the distance along the selected radial direction from the cell boundary. During post-processing, polygons are reconstructed based on these distance maps, and a non-maximum suppression (NMS) algorithm is applied to eliminate redundant polygons.
CPN \citep{upschulte2022contour} uses the same dense probability map to detect individual cells but takes a different approach for proposing cell shapes. CPN regresses the parameters of a cell contour encoded in the frequency domain for each pixel within a cell instance. This method anchors the description of a complete instance into single pixels, compelling the model to establish intrinsic spatial relationships between entire objects and their parts, promoting robust representations with strong generalization. When boundaries are poorly defined or invisible, CPN leverages learned boundary shapes to find plausible separations. However, because the shape modeling acts as a regularizer, an additional module to regress a refinement vector field is necessary to match the precision of dense pixel-based methods in delineating object boundaries. Moreover, as in StarDist, an NMS algorithm is needed to suppress redundant detections.

It is important to highlight that the methods discussed here can be categorized as \textit{one-stage} approaches, where a DL network is followed by a post-processing step. In contrast, a \textit{two-stage} framework involves an initial detection network that identifies cells and generates bounding box predictions, followed by a second network that achieves fine-grained single-cell segmentations. For example, \citet{yi2019attentive} utilize residual network modules \citep{he2016deep} to extract multi-level image features, which are then fused and input into a single-shot multi-box detector \citep{liu2016ssd} for bounding box predictions. The extracted features, together with the bounding box predictions, are fed into a decoder for cell segmentation in a U-Net-like fashion. The well-established Mask-RCNN \citep{he2017mask} remains a benchmark for two-stage approaches, starting with a network that identifies regions likely to contain objects and generating bounding box proposals. In the second stage, object segmentation is accomplished through bounding box regression, along with semantic segmentation via a fully convolutional network. Nevertheless, several one-stage methods have demonstrated superior performance \citep{graham2019hover, stringer2021cellpose, kiran2022denseres, upschulte2022contour, horst2024cellvit}.
\begin{figure*}
\includegraphics[width=\textwidth]{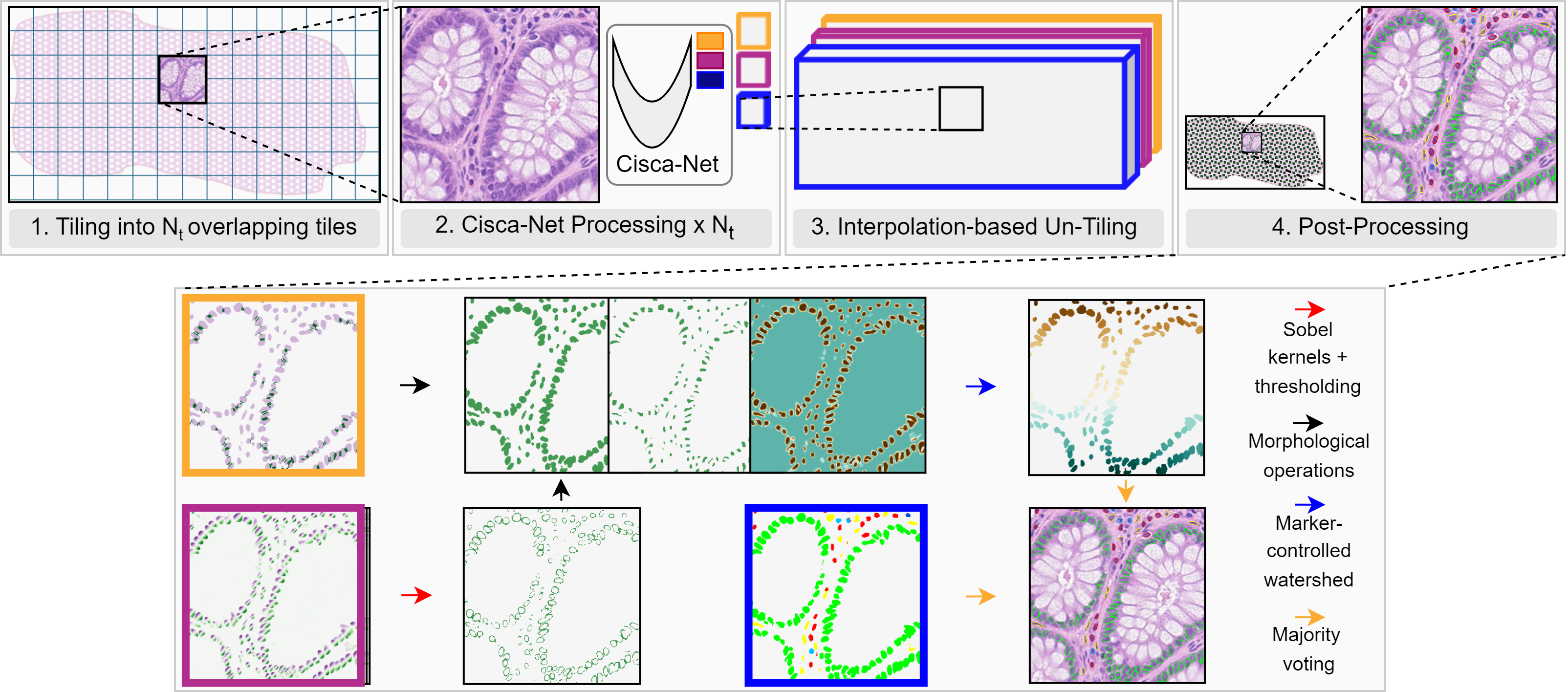}
\caption{An overview of the overall CISCA inference pipeline for cell instance segmentation and classification on WSIs. An image of any size is divided into overlapping patches. Each patch is processed individually by CISCA-Net. The outputs are merged during an untiling step. Finally, a tailored post-processing is applied to generate the instance segmentation map and assign a cell type to each detected cell (cf. Section \ref{sec:post}).} 
\label{CISCdiagram}
\end{figure*}

While the previous methods generally lack the capability for \textit{cell type classification}, they can potentially be enhanced by incorporating an additional head or full-fledged decoder. This allows to simultaneously delineate individual cells and subtype them by predicting an extra semantic segmentation map, where pixel classes coincide with cell classes (e.g., tumoral vs. non-tumoral cells) \citep{graham2023one}. This approach was used in Hover-Net \citep{graham2019hover} and CellViT \citet{horst2024cellvit} or to upgrade StarDist \citep{weigert2022nuclei}. Once the instance segmentation map is established and individual cells are delineated, each cell can be assigned the type that corresponds to the majority of pixels in the predicted cell type semantic segmentation map.

\section{Method}
\label{method}

We propose \textit{CISCA} as a pipeline for the automatic segmentation and classification of individual cells in histological images. As depicted in Fig. \ref{CISCdiagram}, a WSI of any size is first divided into $256 \times 256$ patches with 50\% overlap. Each patch is processed individually by the deep learning model \textit{CISCA-Net}. The outputs from the model are then merged during an untiling step. Subsequently, a series of post-processing steps are adopted to produce the instance segmentation map and assign a class to each detected cell (cf. Section \ref{sec:post}).

\subsection{CISCA-Net}
\subsubsection{Architecture}
\begin{figure*}
\includegraphics[width=\textwidth]{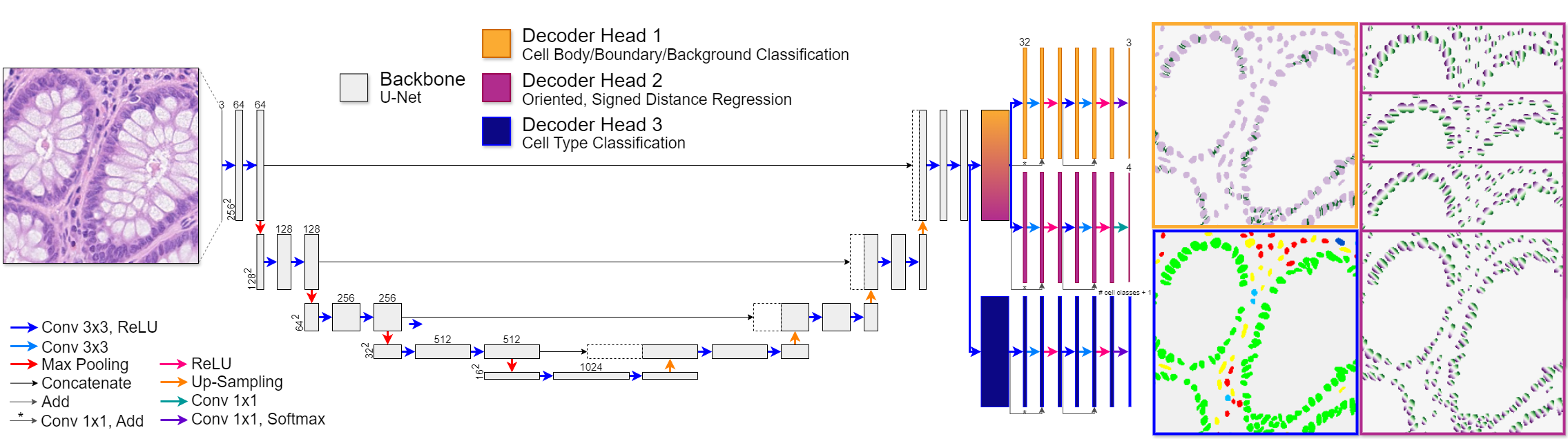}
\caption{A diagram of the architecture of the suggested CISCA-Net for cell instance segmentation and classification. CISCA-Net is a DL architecture featuring a lightweight U-Net backbone and three convolutional heads. The first head (yellow) focuses on classifying pixels into boundary between closely located cells (green), cell bodies (pink), and background (white). The second head (purple) handles the regression of four oriented distance maps, predicting distances from the cell centroid for each pixel. Distance maps are normalized to increase from a minimum of $-1$ (purple) to a maximum of $1$ (green). The third head (blue) classifies pixels into different cell types, depending on the dataset.} \label{CISCArchitecture}
\end{figure*}

CISCA's core component is CISCA-Net, a relatively lightweight convolutional network with a U-Net backbone \citep{ronneberger2015u} of depth four and three heads, as shown in Fig. \ref{CISCArchitecture}. The first two heads focus on cell instance segmentation, while the third handles cell classification. CISCA-Net tackles the cell instance segmentation problem using a multi-task approach that integrates pixel classification (semantic segmentation) and regression. The first head classifies each pixel to either cell bodies (CB), BG, or inter-cell boundary (BD), as shown in Fig. \ref{40xPadova} d) and Fig. \ref{CISCArchitecture}. The second head generates four distance maps, represented as a four-channel tensor. Each channel corresponds to a distance map along a specific direction: vertical, horizontal, diagonal from top left to bottom right, and diagonal from bottom left to top right. In each distance map, every pixel value within a cell indicates the distance from the line through the cell's centroid that is orthogonal to the distance map's direction. For example, the horizontal distance map measures the distance from the vertical line. Distance maps are normalized to lie in the range $[-1, 1]$, as shown in Fig. \ref{CISCArchitecture}, where each cell is depicted by a color gradient, with pixel values increasing from a minimum of $-1$ (purple) to a maximum of $1$ (green). The value $0$ corresponds to the `$0-line$' through a cell centroid. The background is also set to $0$.
The third head classifies pixels into cell types. The input to CISCA-Net's heads is obtained by processing the output of the shared U-Net backbone via two parallel $3 \times 3$ convolution layers yielding two separate feature maps with dimensions $256 \times 256 \times 256$. The first feature map is shared by the first two heads, while the second feature map is dedicated to cell classification. Each head processes its feature map via two residual blocks followed by a $1 \times 1$ convolution to generate tensors with 3, 4, and $T$ channels, respectively, where $T$ is the number of cell types. The first and third head are completed by a softmax layer. The third head is absent when classification into cell types is not of interest. 

Our network design is primarily inspired by Hover-Net \citep{graham2019hover}, StarDist \citep{schmidt2018cell}, and \citet{wu2022general}. We incorporate two diagonal distance maps into the predictions of Hover-Net. Since distance maps are ultimately post-processed to determine boundary predictions for separating distinct cells, we hypothesized that regressing diagonal distance maps, along with horizontal and vertical ones, would yield more robust predictions. This approach has a parallel in StarDist, which regresses distances of pixels from the cell centroid along radial directions to construct star-convex polygons. Unlike StarDist, which uses at least 64 distance maps for competitive performance, our approach employs only four distance maps. Additionally, we depart from the Hover-Net architecture, which employs a Preact-ResNet50 backbone along with two independent decoders. Instead, we implement a simpler U-Net backbone based on the original U-Net design, followed by more straightforward convolutional heads, resulting in a significant reduction in the number of parameters. Finally, we assume that more precise cell separation can be achieved by classifying pixels into BD, CB, and BG, rather than solely performing foreground/background separation as done in Hover-Net. To achieve accurate predictions, both the pixel classification and regression heads must concentrate on localizing the boundaries between closely positioned cells, facilitating more effective feature sharing. Our approach is related to \citet{wu2022general}, where boundaries are defined as the portions of cell contours shared between \textit{touching} cells. We argue that this definition of the boundary class is tricky to learn due to the typical absence of sharp edges between cells in close proximity. Consequently, the touching/not touching distinction as defined by the annotator(s) has a random component that ultimately leads to inconsistent GT and, we argue, more uncertain predictions. Inspired by \citet{pena2020j}, we define the boundary class as the regions of cell bodies between closely positioned cells, forming a superset of the boundaries between touching cells. This leads to a more consistent GT and a reduction of class imbalance, since the proportion of pixels belonging to BD, which is typically underrepresented compared to CB and BG, increases.

\subsubsection{Loss function}
\label{sec:losses}
CISCA-Net is trained using a composite loss function $\mathcal{L} = \mathcal{L}^{p} + \mathcal{L}^{r} + \mathcal{L}^{t}$, which combines terms for pixel classification into BD, CB, BG ($\mathcal{L}^{p}$), distance maps regression ($\mathcal{L}^{r}$), and pixel classification into cell types ($\mathcal{L}^{t}$). In the following, we refer to the heads' output tensors as $\hat{\mathbf{Y}}^{p} = [\hat{\mathbf{Y}}^{p}_{ijk}] \in [0,1]^{N \times N \times P}$, $\hat{\mathbf{Y}}^{r} =[\hat{\mathbf{Y}}^{r}_{ijk}] \in \mathbb{R}^{N \times N \times R}$, and $\hat{\mathbf{Y}}^{t} = [\hat{\mathbf{Y}}^{t}_{ijk}] \in [0,1]^{N \times N \times T}$, respectively, where $P=3$, $R=4$, and $T>1$ are the number of pixel classes, distance maps, and cell classes, respectively, $N$ matches the width/height of the input image, and $\hat{\mathbf{Y}}^{*}_{ijk}$ indicates a tensor element. The ground truths (GT) are represented as $\mathbf{Y}^{p}$, $\mathbf{Y}^{r}$, and $\mathbf{Y}^{t}$. 

The first loss component $\mathcal{L}^{p}$ is given by:
\begin{equation}
\mathcal{L}^{p} = \lambda_1\mathcal{C}^{p} + \lambda_2\mathcal{D}_1 + \lambda_3\mathcal{D}_2 
\label{pcloss}
\end{equation}
where $\lambda_1 = \lambda_3 = 2$ and $\lambda_2 = 1$ are weighting factors set via hyperparameter validation \citep{vadori2023ncis}. $\mathcal{C}^{p}$ is the categorical cross-entropy between $\mathbf{Y}^{p}$ and $\hat{\mathbf{Y}}^{p}$:
\begin{equation}
\mathcal{C}^{p} = -\frac{1}{N^2}\sum_{i,j=1}^{N}\sum_{k=1}^{P}\mathbf{Y}^{p}_{ijk}\log(\hat{\mathbf{Y}}^{p}_{ijk})
\end{equation}
$\mathcal{D}_k$ is the Dice loss for class $k$, with $k=1$ corresponding to class BD and $k=2$ corresponding to class CB:
 \begin{equation}
\mathcal{D}_k = 1 - \frac{2\sum_{i,j=1}^{N} (\mathbf{Y}^{p}_{ijk} \odot \hat{\mathbf{Y}}^{p}_{ijk})}{\sum_{i,j=1}^{N} \mathbf{Y}^{p}_{ijk} + \sum_{i,j=1}^{N} \hat{\mathbf{Y}}^{p}_{ijk}}
\end{equation}
$\mathcal{L}^{p}$ is inspired by \citet{wu2022general}, where $\mathcal{C}$ is combined with Dice losses for the minority classes to encourage correct classification of pixels from the BD and CB classes. Note, however, that in \citet{wu2022general}, the weighting factors differ, and the Dice losses are replaced by \textit{soft} Dice losses.
The second loss component $\mathcal{L}^{r}$ is given by:
\begin{equation}
\mathcal{L}^{r} = \lambda_4\mathcal{A} + \lambda_5\mathcal{S}
\label{pcloss2}
\end{equation}
where $\lambda_4 = \lambda_5 = 2$ are weighting factors.
Following \citet{graham2019hover}, $\mathcal{A}$ and $\mathcal{S}$ are defined as follows. $\mathcal{A}$ represents a normalized mean absolute error between $\mathbf{Y}^{r}$ and $\hat{\mathbf{Y}}^{r}$. While \citet{graham2019hover} consider contributions from regions corresponding to cell bodies, we use a mask to give greater weight to regions in and \textit{around} cell bodies and less weight to BG pixels:
\begin{equation}
\mathcal{A} = \frac{\sum_{i,j=1}^{N}\sum_{k=1}^{R}|\mathbf{Y}^{r}_{ijk} - \hat{\mathbf{Y}}^{r}_{ijk}| \odot \mathbf{Y}^{\mathit{mask}}_{ij}}{\sum_{i,j=1}^{N}\sum_{k=1}^{R} \mathbf{Y}^{\mathit{mask}}_{ij}}
\end{equation}
where $R$ is the number of distance maps, and $\mathbf{Y}^{\mathit{mask}} = [\mathbf{Y}^{\mathit{mask}}_{ij}]$ is a binary mask obtained via summing a morphological dilation of $[\mathbf{Y}^{p}_{ij1} + \mathbf{Y}^{p}_{ij2}]$ and $0.05\cdot [\mathbf{Y}^{p}_{ij3}]$. $\mathcal{S}$ is defined as:
 \begin{equation}
\mathcal{S} = \frac{\sum_{i,j=1}^{N}\sum_{k=1}^{R}(\nabla_k \mathbf{Y}^{r}_{ijk} - \nabla_k \hat{\mathbf{Y}}^{r}_{ijk})^2 \odot \mathbf{Y}^{\mathit{mask}}_{ij}}{\sum_{i,j=1}^{N}\sum_{k=1}^{R} \mathbf{Y}^{\mathit{mask}}_{ij}}
\end{equation}
where the $\nabla$ operator computes the finite derivative by convolving a map with a $5 \times 5$ Sobel kernel, oriented and directed according to the distance map, the same used in the post-processing phase. $\mathcal{S}$ essentially enforces the gradient of the distance maps to be aligned with the gradients in the GT. 

Finally, the loss component $\mathcal{L}^{t}$ is given by:
\begin{equation}
\mathcal{L}^{t} = \lambda_6 \cdot (\mathcal{C}^{t} + \mathcal{T})
\label{ccloss}
\end{equation}
where $\lambda_6 = 5$, and $\mathcal{C}^{t}$ is the categorical cross-entropy between $\mathbf{Y}^{t}$ and $\hat{\mathbf{Y}}^{t}$, with contributions weighted by $\mathbf{Y}^{\mathit{mask}}_{ij}$ as described previously. $\mathcal{T}$ is a pixel-wise formulation of the \textit{Tversky loss} \citep{salehi2017tversky} computed as:
\begin{equation}
\mathcal{T} = \frac{1}{N^2 \cdot T} \sum_{i,j=1}^{N} \sum_{k=1}^{T}  \mathcal{T}_{ijk}
\end{equation}
where 
\begin{equation}
\mathcal{T}_{ijk} = 1 - \frac{\mathbf{Y}^{t}_{ijk}\! \odot\! \hat{\mathbf{Y}}^{t}_{ijk}}{\mathbf{Y}^{t}_{ijk} \!\odot \!\hat{\mathbf{Y}}^{t}_{ijk} \!+\! \alpha \mathbf{Y}^{t}_{ijk} (1\! - \!\hat{\mathbf{Y}}^{t}_{ijk})\! +\! (1 \!-\! \alpha) (1\! -\! \mathbf{Y}^{t}_{ijk}) \hat{\mathbf{Y}}^{t}_{ijk}}
\end{equation}
With $\alpha > 0.5$, a higher penalty is applied to false negatives for a given class, which helps in scenarios with class imbalance where pixels from a minority class are misclassified.


\subsubsection{Ground Truth}
\label{sec:GT}
The GT for distance maps and pixel classification is derived from the GT label map. Distance maps are generated by modifying the approach described in \citet{graham2019hover} to incorporate diagonals. For pixel classification, a custom algorithm produces a map $\mathbf{Y}^{p,0} = [\mathbf{Y}^{p,0}_{ij}] \in [1,2,3]^{N \times N}$ that categorizes pixels as BD (1), CB (2), or BG (3). This algorithm initializes an empty matrix $\mathbf{T}$ and processes a binary mask for each instance by dilating it and adding it to $\mathbf{T}$. Pixels where $\mathbf{T}_{ij} > 1$ denote areas between closely positioned cells, which are further expanded by two sequential dilations. Within cell instances, as per the GT label map, these expanded areas are marked as BD in $\mathbf{Y}^{p,0}$, while other pixels are classified as CB. Pixels outside cell instances are labeled as BG. Finally, morphological operations are applied to remove small residual components for each class. As specified in \ref{sec:losses}, the actual GT map used for training  CISCA-Net is a tensor $\mathbf{Y}^{p} = [\mathbf{Y}^{p}_{ijk}] \in [0,1]^{N \times N \times 3}$, which is derived by one-hot encoding $\mathbf{Y}^{p,0}$.

\subsubsection{Oversampling} \label{sec:Oversampling} 
Cell classes in CoNIC and PanNuke are imbalanced, as shown in Tables \ref{tab:cellclasscounts} and \ref{tab:cellclasscounts2}. For instance, epithelial cells constitute approximately $50\%$ of the cells in the CoNIC training data, whereas neutrophils and eosinophils make up around $1\%$. In PanNuke, neoplastic cells represent the majority at $40.2\%$, while dead cells account for just $1.4\%$. To address class imbalance, we applied oversampling. 
For each minority class (i.e., all classes except the majority class), we oversampled the training set by adding a number of images calculated  for each class $t$ as $N_{extra, \ t}= \alpha_t \cdot \frac{\beta_t}{\max_t \beta_t} \cdot N_{train}$, where $N_{train}$ denotes the total number of images in the training set and $\alpha_t$ and $\beta_t$ are multiplicative factors. Specifically, $\beta_t = \sqrt{\frac{C_{train}}{C_t}}$, with $C_{train}$ being the total number of cells in the training set and $C_t$ the number of cells in class $t$, following the approach of \citet{weigert2022nuclei}. $\alpha_t$ can be set to $1$ for all classes. We did so PanNuke, while for CoNIC, we chose $\alpha_t$ values of $1.6$ for neutrophils and eosinophils, and $0.9$ for other cell types to further increase the relative number of images sampled for neutrophils and eosinophils.
For each minority cell class, images added to the training set were sampled with a probability proportional to the ratio of the number of cells from the class in the image to the total number of cells from the class across all images. This approach ensured that images containing a higher number of minority class cells had a greater chance of being selected. The original and oversampled image and cell counts are reported in Table \ref{tab:split_composition}. For CoNIC, the training set increased from $3,589$ images to $16,690$; for PanNuke, from $5,688$ to $16,035$. The updated proportions are shown in Tables \ref{tab:cellclasscounts} and \ref{tab:cellclasscounts2}. The proportion of epithelial cells in CoNIC decreased to $36.7\%$, while neutrophils and eosinophils increased to $2.7\%$ and $1.8\%$, respectively. In PanNuke, neoplastic cells were reduced to $34.6\%$, and dead cells increased to $14.6\%$. 
\subsubsection{Stain Augmentation} \label{sec:StainNormalizationAugmentation} 
\begin{figure*}
\includegraphics[width=\textwidth]{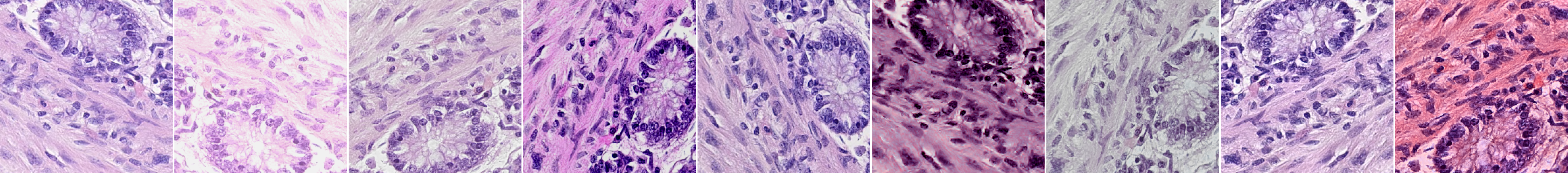}
\includegraphics[width=\textwidth]{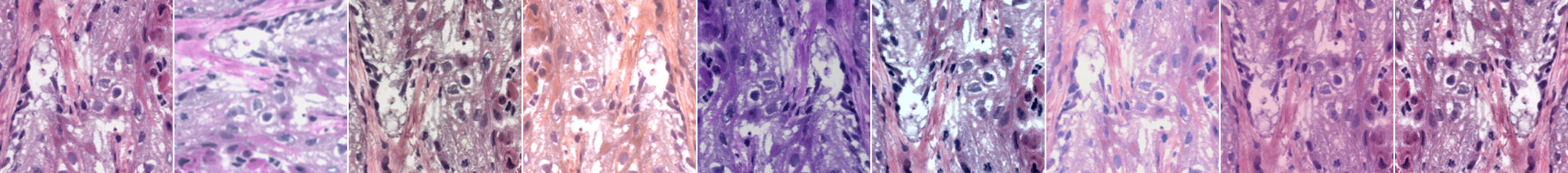}
\includegraphics[width=\textwidth]{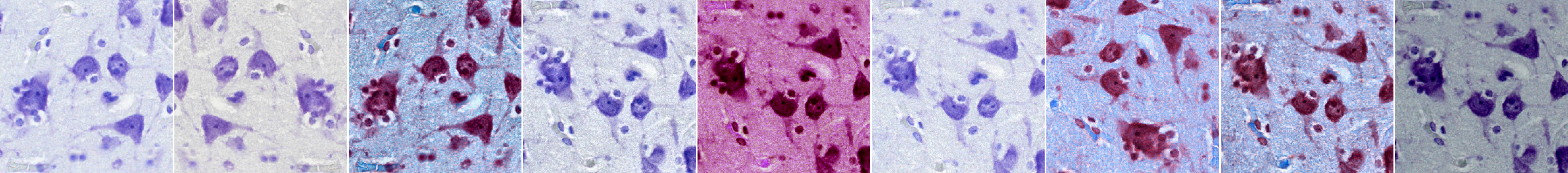}
\includegraphics[width=\textwidth]{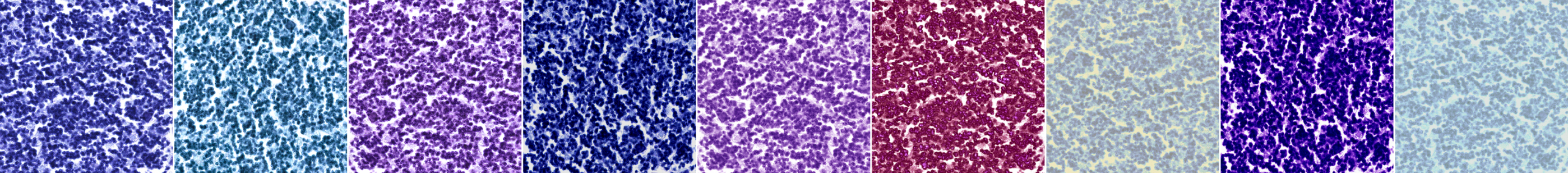}
\caption{Effect of the augmentation strategy on sample patches from the CoNIC (first row), PanNuke (second row), CytoDArk0\_20x\_256 (third row), and CytoDArk0\_40x\_1024 (fourth row) datasets. The original patch is shown in the leftmost position of a row and is followed on the right by patches obtained by applying random transformations as detailed in Section \protect\ref{sec:StainNormalizationAugmentation}.} \label{fig:StainNormalizationAugmentation}
\end{figure*}

Stain shifts are common in histology, due to the variability of staining protocols, staining scanners, manufacturers, batches of staining, which eventually hinders a model performance at inference time \citep{vasiljevic2021towards}. To desensitize the learned model to stain variations and improve generalization, we adopted a stain augmentation (SA) technique. SA aims to broaden the range of stain styles by simulating staining variations while maintaining morphological features. During training, we randomly apply one of the following SA strategies.
a) Apply the method proposed by \citep{shen2022randstainna}, which defines the stain style of an image using the averages and standard deviations of each channel in the LAB color space. These parameters are computed prior to training, and their average and standard deviation parameterize two Gaussian distributions for the average stain intensity and stain standard deviation. During training, a random sample is drawn from these distributions to generate a stain style, which is then applied to the source image via linear transformations in LAB space before converting it to RGB.
b) Apply random modifications to brightness, contrast, or saturation.
c) Adjust the stain of the source image to match the appearance of one of a set of fixed template images using a normalization method randomly chosen from \citet{ruifrok2001quantification}, \citet{macenko2009method}, \citet{vahadane2016structure}, or \citet{reinhard2001color}.
The effect of the augmentation strategy on sample patches is shown in Fig. \ref{fig:StainNormalizationAugmentation}.
\subsubsection{Training} 
\label{sec:Training} 
Training is performed for up to $300$ epochs with a batch size of $N_{batch} = 4$. We use the AMSGrad optimizer with a clip norm of $0.001$ and an initial learning rate of $5 \times 10^{-4}$, incorporating step decay (the learning rate is halved every $10$ epochs if the validation loss does not decrease) and early stopping with a patience of $20$ epochs. 
During each epoch, a set of $\left\lceil \frac{N_{train}}{N_{batch}}  \times \left( \frac{H}{256} \right)^2 \right\rceil$ batches is created by sampling images with replacement from the oversampled training set, where $N_{train}$ is the number of images in the oversampled training set, and $H \ge 256$ is the height of the original (uncropped) images. A series of random augmentations are chosen from rotations, flips, elastic deformations, stain augmentation (cf. Section \ref{sec:StainNormalizationAugmentation}), Gaussian noise, blurring, and sharpening and applied to each image and GT masks when needed. If $H > 256$, images are cropped to size $256\times256$.
\subsection{Untiling and Post-Processing}
\label{sec:post}
For WSIs, $256\times256$ tile predictions are blended together via interpolation with a second order spline window function that weights pixels when merging patches \citep{guillaume2017}. According to the notation in Section \ref{sec:losses}, this yields predictions $\hat{{\mathbf{Y}}^{p}}$, $\hat{\mathbf{Y}}^{r}$, $\hat{\mathbf{Y}}^{t}$ for the entire image, which are post-processed as illustrated in Fig. \ref{CISCdiagram} and described below.  

A binary mask $\mathbf{E}$ is defined for cell edges by setting pixels as $\mathbf{E}_{ij} = \mathbb{I} (max(\nabla_1\hat{\mathbf{Y}}^{r}_{ij1},\nabla_2\hat{\mathbf{Y}}^{r}_{ij2},\nabla_3\hat{\mathbf{Y}}^{r}_{ij3},\nabla_4\hat{\mathbf{Y}}^{r}_{ij4}) > \theta_1)$, where each distance map $[\hat{\mathbf{Y}}^{r}_{ijk}],  k= 1,...,4,$ is normalized between $0$ and $1$, $\nabla_k,  k= 1,...,4,$ represents the convolution with of a Sobel kernel whose direction matches that of the distance map, and $\theta_1=0.57$ is a binarization threshold. $\theta_1$ controls the amount of pixels considered as edges. If set too low, the edges encroach into the cell bodies, while if set too high, the edges are not adequately demarcated. The optimal value $\theta_1=0.57$ was set via hyperparameter validation. Note that an analogous threshold in Hover-Net is set to $\theta_1=0.4$.
The marker-controlled watershed labelling is then applied to label pixels that belong to a foreground binary mask representing pixels from BD or CB classes ($\hat{\mathbf{Y}}^{p}_{ij1}+\hat{\mathbf{Y}}^{p}_{ij2}>\hat{\mathbf{Y}}^{p}_{ij3}$).
The markers are defined by a binary mask $\mathbf{B}$ that corresponds to the CB class refined by $\mathbf{E}$ ($ \hat{\mathbf{Y}}^{p}_{ij2}>\hat{\mathbf{Y}}^{p}_{ij1}$ and $\hat{\mathbf{Y}}^{p}_{ij2}>\hat{\mathbf{Y}}^{p}_{ij3}$ and $\mathbf{E}_{ij}= 0$). The topological map for watershed is derived from the complement of $[\hat{\mathbf{Y}}^{p}_{ik2}]$ summed to the complement of the product of the foreground binary mask and the complement of $\mathbf{E}$. This results in the instance segmentation map (or label map) $\hat{\mathbf{Y}}^{l}$, where the area of each identified cell is assigned a unique positive integer label. Cells smaller than a threshold $\theta_2$, which can be determined based on the user's knowledge of the minimum cell dimensions, are deleted. Subsequently, using $\hat{\mathbf{Y}^{c}}$, each detected cell is classified based on the cell type corresponding to the majority of pixels within the cell instance. To be noted that the morphological structuring elements used for the post-processing process are specifically chosen according to the image magnification (20x or 40x).
\subsection{Implementation}
CISCA is implented on Python. The source code is available at \url{https://github.com/vadori/cytoArk}. CISCA-Net is currently implemented on the Tensorflow framework. Training was conducted on an NVIDIA Quadro RTX 6000 with $24$ GB of memory, and the experiments were performed on a system running Windows 10 with $128$ GB of RAM.

\begin{table*}[]
\captionsetup{aboveskip=2pt, belowskip=1pt}
\caption{Number of images and cells for each dataset used for training in this study. Data is detailed for each split (training, validation, and testing) along with the percentages relative to the total. For CoNIC and PanNuke, which include cell classes, oversampling was applied to address class imbalance, as detailed in Section \ref{sec:StainNormalizationAugmentation}. Therefore, numbers are reported both without and with oversampling (w/o overs. and w/ overs., respectively).}
\label{tab:split_composition}
\centering
\begingroup
\setlength{\tabcolsep}{1.6pt}
\begin{tabular}{@{}lllllllllllll@{}}
\toprule
\multicolumn{1}{c}{} &
   &
  \multicolumn{5}{c}{\textbf{Image   Count}} &
   &
 \multicolumn{5}{c}{\textbf{Cell   Count}} \\ \cmidrule(lr){3-7} \cmidrule(l){9-13} 
 &
   &
  Training &
  Validation &
  Testing &
   &
  Total &
   &
  Training &
  Validation &
  Testing &
   &
  Total \\
{ CoNIC w/o overs.} &
   &
  3,589 \scriptsize{(72.1\%)} &
  464 \scriptsize{(9.3\%)} &
  928 \scriptsize{(18.6\%)} &
   &
  4,981 &
   &
  412,456 \scriptsize{(72.3\%)} &
  56,331 \scriptsize{(9.9\%)} &
  101,808 \scriptsize{(17.8\%)} &
   &
  570,595 \\
{ CoNIC w/ overs.} &
   &
  16,690 \scriptsize{(92.3\%)} &
  464 \scriptsize{(2.6\%)} &
  928 \scriptsize{(5.1\%)} &
   &
  18,082 &
   &
  2,338,886 \scriptsize{(93.7\%)} &
  56,331 \scriptsize{(2.3\%)} &
  101,808 \scriptsize{(4.1\%)} &
   &
  2,497,025 \\
{ PanNuke w/o overs.} &
   &
  5,688 \scriptsize{(72.0\%)} &
  632 \scriptsize{(8.0\%)} &
  1,581 \scriptsize{(20.0\%)} &
   &
  7,901 &
   &
  137,859 \scriptsize{(71.9\%)} &
  15,397 \scriptsize{(8.0\%)} &
  38,545 \scriptsize{(20.1\%)} &
   &
  191,801 \\
{ PanNuke w/ overs.} &
   &
  16,035 \scriptsize{(87.9\%)} &
  632 \scriptsize{(3.5\%)} &
  1,581 \scriptsize{(8.7\%)} &
   &
  18,248 &
   &
  594,432 \scriptsize{(91.7\%)} &
  15,397 \scriptsize{(2.4\%)} &
  38,545 \scriptsize{(5.9\%)} &
   &
  648,374 \\
{ CytoDArk0\_20x\_1024} &
   &
  45 \scriptsize{(65.2\%)} &
  6 \scriptsize{(8.7\%)} &
  18 \scriptsize{(26.1\%)} &
   &
  69 &
   &
  23,807 \scriptsize{(61.4\%)} &
  5,414 \scriptsize{(14.0\%)} &
  9,534 \scriptsize{(24.6\%)} &
   &
  38,755 \\
{ CytoDArk0\_40x\_1024} &
   &
  172 \scriptsize{(74.1\%)} &
  24 \scriptsize{(10.3\%)} &
  36 \scriptsize{(15.5\%)} &
   &
  232 &
   &
  23,367 \scriptsize{(64.3\%)} &
  5,579 \scriptsize{(15.3\%)} &
  7,412 \scriptsize{(20.4\%)} &
   &
  36,358 \\
{ CytoDArk0\_20x\_256} &
   &
  720 \scriptsize{(65.2\%)} &
  96 \scriptsize{(8.7\%)} &
  288 \scriptsize{(26.1\%)} &
   &
  1,104 &
   &
  26,182 \scriptsize{(61.6\%)} &
  5,891 \scriptsize{(13.9\%)} &
  10,459 \scriptsize{(24.6\%)} &
   &
  42,532 \\
{ CytoDArk0\_40x\_256} &
   &
  2,752 \scriptsize{(74.1\%)} &
  384 \scriptsize{(10.3\%)} &
  576 \scriptsize{(15.5\%)} &
   &
  3,712 &
   &
  { 28,056 \scriptsize{(64.8\%)}} &
  6,613 \scriptsize{(15.3\%)} &
  8,609 \scriptsize{(19.9\%)} &
   &
  43,278 \\ \bottomrule
\end{tabular}
\endgroup

\end{table*}

\begin{table*}[]
\captionsetup{aboveskip=2pt, belowskip=1pt}
\caption{Number of cells for each cell class in CoNIC and PanNuke. Data is detailed for each split. \textit{Training w/o overs.} represents the number of cells before applying the oversampling strategy. \textit{Training w/ overs.} represents the number of cells after applying it. }
\label{tab:cellclasscounts}
\centering
\begin{tabular}{@{}lllllll@{}}
\toprule
 &
  \multicolumn{6}{c}{\textbf{CoNIC}} \\ \cmidrule(l){2-7} 
 &
  Neutrophil &
  Epithelial &
  Lymphocyte &
  Plasma &
  Eosinophil &
  Connective \\
Training w/o overs. &
  4,232 \scriptsize{(1.0\%)} &
  \multicolumn{1}{r}{202,125 \scriptsize{(49.0\%)}} &
  85,977 \scriptsize{(20.8\%)} &
  23,171 \scriptsize{(5.6\%)} &
  3,156 \scriptsize{(0.8\%)} &
  93,795 \scriptsize{(22.7\%)} \\
Training w/ overs. &
  62,327 \scriptsize{(2.7\%)} &
  858,938 \scriptsize{(36.7\%)} &
  614,564 \scriptsize{(26.3\%)} &
  172,840 \scriptsize{(7.4\%)} &
  41,543 \scriptsize{(1.8\%)} &
  588,674 \scriptsize{(25.2\%)} \\
Validation &
  237 \scriptsize{(0.4\%)} &
  25,139 \scriptsize{(44.6\%)} &
  15,250 \scriptsize{(27.1\%)} &
  3,850 \scriptsize{(6.8\%)} &
  196 \scriptsize{(0.3\%)} &
  11,659 \scriptsize{(20.7\%)} \\
Testing &
  617 \scriptsize{(0.6\%)} &
  55,307 \scriptsize{(54.3\%)} &
  19,732 \scriptsize{(19.4\%)} &
  4,959 \scriptsize{(4.9\%)} &
  500 \scriptsize{(0.5\%)} &
  20,693 \scriptsize{(20.3\%)} \\ \bottomrule
\end{tabular}
\end{table*}
\begin{table*}[]
\captionsetup{aboveskip=2pt, belowskip=1pt}
\caption{Number of cells for each cell class in CoNIC and PanNuke. Data is detailed for each split. \textit{Training w/o overs.} represents the number of cells before applying the oversampling strategy. \textit{Training w/ overs.} represents the number of cells after applying it. }
\label{tab:cellclasscounts2}
\centering
\begin{tabular}{@{}llllll@{}}
\toprule
           & \multicolumn{5}{c}{\textbf{PanNuke}}                                                                                                               \\ \cmidrule(l){2-6} 
           & Neoplastic                   & Inflammatory                & Connective                   & Dead                     & Non-neoplastic/Epithelial   \\
Training w/o overs. &
  55,478 \scriptsize{(40.2\%)} &
  23,342 \scriptsize{(16.9\%)} &
  37,856 \scriptsize{(27.5\%)} &
  1,882 \scriptsize{(1.4\%)} &
  19,301 \scriptsize{(14.0\%)} \\
Training w/ overs. &
  205,653 \scriptsize{(34.6\%)} &
  106,793 \scriptsize{(18.0\%)} &
  122,179 \scriptsize{(20.6\%)} &
  86,677 \scriptsize{(14.6\%)} &
  73,130 \scriptsize{(12.3\%)} \\
Validation & 6,311 \scriptsize{(41.0\%)}  & 2,575 \scriptsize{(16.7\%)} & 3,963 \scriptsize{(25.7\%)}  & 341 \scriptsize{(2.2\%)} & 2,207 \scriptsize{(14.3\%)} \\
Testing    & 15,742 \scriptsize{(40.8\%)} & 6,670 \scriptsize{(17.3\%)} & 10,071 \scriptsize{(26.1\%)} & 725 \scriptsize{(1.9\%)} & 5,337 \scriptsize{(13.8\%)} \\ \bottomrule
\end{tabular}
\end{table*}

\section{Datasets}
\begin{table*}[]
\captionsetup{aboveskip=4pt, belowskip=2pt}
\caption{Number of images and cells for each brain region and species in CytoDArk0\_20x\_1024 and CytoDArk0\_40x\_2048 data(sub)sets, part of a new dataset called \textit{CytoDArk0}. CE: \textit{Cerebellum}. HI: \textit{Hippocampus}. AC: \textit{Auditory Cortex}. VC: \textit{Visual Cortex}.}
\label{tab:CytoDArksummary}
\centering
\begingroup
\setlength{\tabcolsep}{4pt}
\begin{tabular}{@{}llcclclclcclcclclc@{}}
\toprule
\multicolumn{18}{c}{\textbf{CytoDArk0}: A novel dataset for cytoarchitecture studies on Nissl-stained histological images of the mammalian brain} \\ \midrule
 &
   &
  \multicolumn{9}{c}{\begin{tabular}[c]{@{}c@{}}\textbf{CytoDArk0\_20x\_1024}\\ \textbf{69} RGB images at \textbf{20x} with size \textbf{1024x1024}\end{tabular}} &
   &
  \multicolumn{6}{c}{\begin{tabular}[c]{@{}c@{}}\textbf{CytoDArk0\_40x\_2048}\\ \textbf{58} RGB images at \textbf{40x} with size \textbf{2048x2048}\end{tabular}} \\ \cmidrule(lr){3-11} \cmidrule(l){13-18} 
Brain Area &
   &
  \multicolumn{2}{c}{\textbf{CE}} &
   &
  \textbf{HI} &
   &
  \textbf{AC} &
   &
  \multicolumn{2}{c}{\textbf{VC}} &
   &
  \multicolumn{2}{c}{\textbf{CE}} &
   &
  \textbf{HI} &
   &
  \textbf{AC} \\ \cmidrule(lr){3-4} \cmidrule(lr){6-6} \cmidrule(lr){8-8} \cmidrule(lr){10-11} \cmidrule(lr){13-14} \cmidrule(lr){16-16} \cmidrule(l){18-18} 
Species &
   &
  Bovine &
  Chimp &
   &
  Mouse &
   &
  B. dolphin &
   &
  Chimp &
  Macaque &
   &
  Bovine &
  Chimp &
   &
  Mouse &
   &
  B. dolphin \\
\#Images &
   &
  1 &
  3 &
   &
  4 &
   &
  53 &
   &
  4 &
  4 &
   &
  1 &
  3 &
   &
  1 &
   &
  53 \\
\#Cells &
   &
  3,680 &
  7,381 &
   &
  3,213 &
   &
  22,804 &
   &
  878 &
  799 &
   &
  3,680 &
  7,385 &
   &
  1,330 &
   &
  22,837 \\ \bottomrule
\end{tabular}
\endgroup

\end{table*}
\begin{figure*}[h]%
\centering
\begin{subfigure}{.5\columnwidth}
\includegraphics[width=\columnwidth]{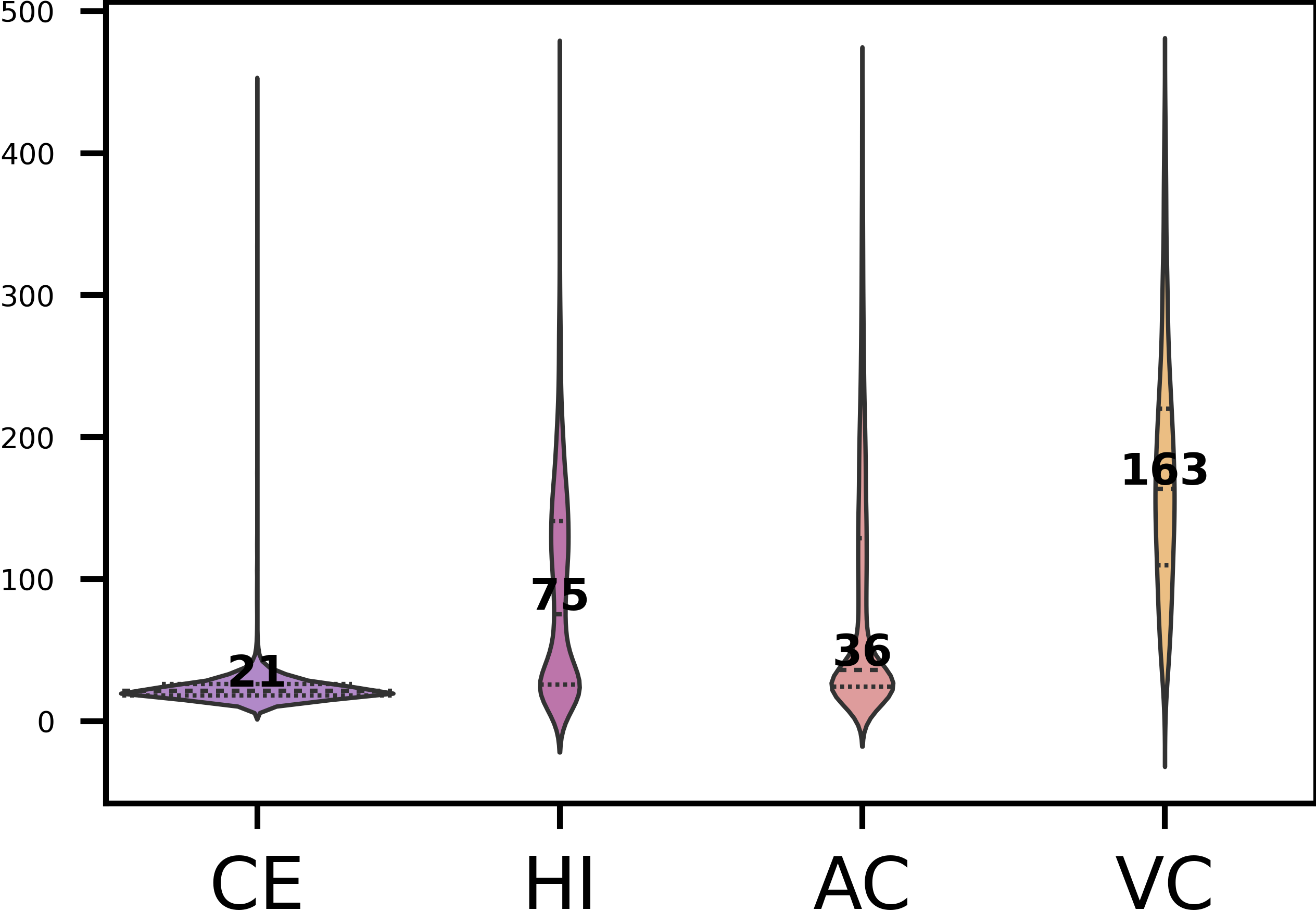}%
\caption{Cell Area ($\mu m^2$)}%
\label{subfiga}%
\end{subfigure}\hfill%
\begin{subfigure}{.5\columnwidth}
\includegraphics[width=\columnwidth]{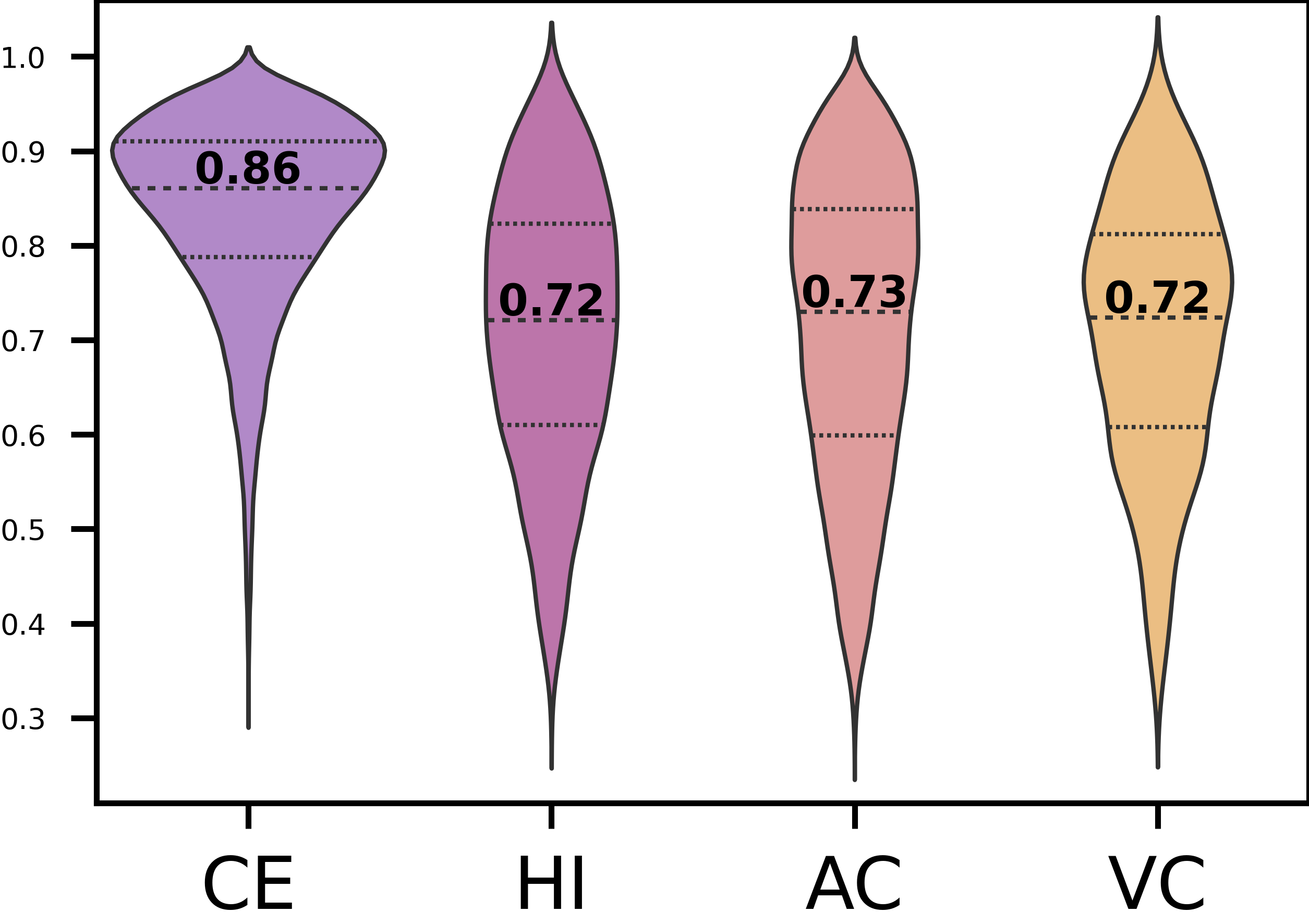}%
\caption{Axis Ratio}%
\end{subfigure}\hfill%
\begin{subfigure}{.5\columnwidth}
\includegraphics[width=\columnwidth]{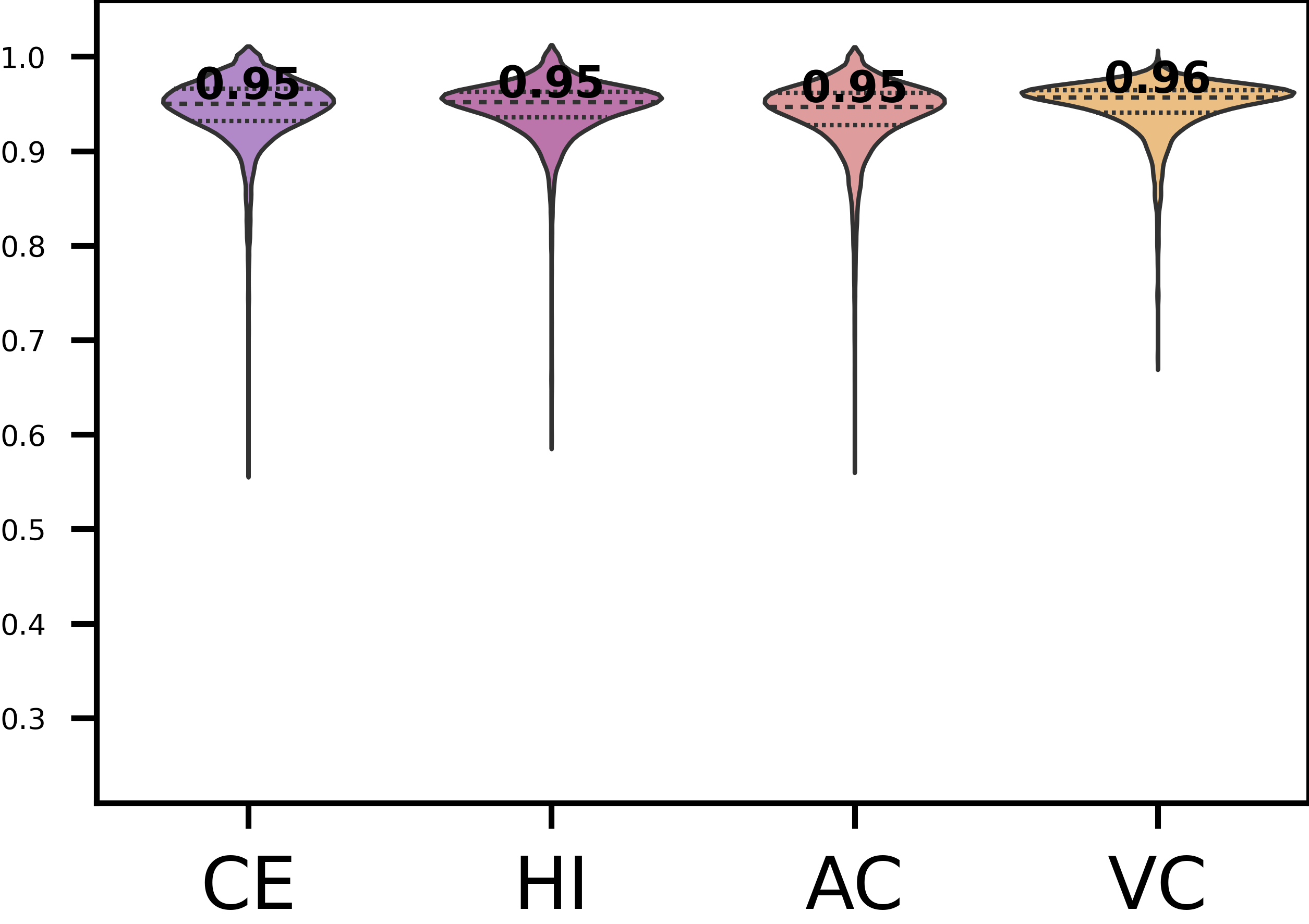}%
\caption{Solidity}%
\end{subfigure}\hfill%
\begin{subfigure}{.5\columnwidth}
\includegraphics[width=\columnwidth]{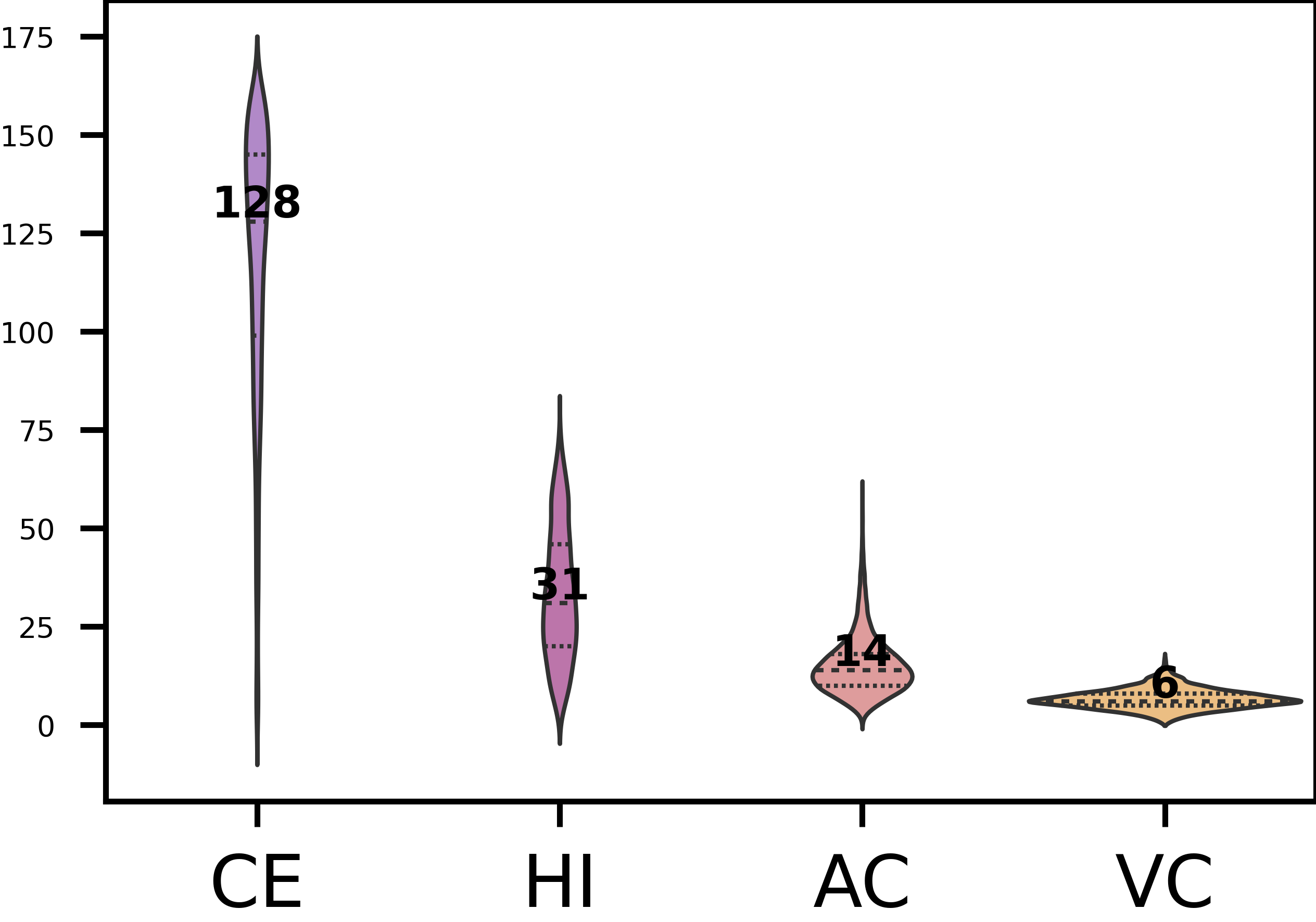}%
\caption{Neighbourhood Density (\#cells)}%
\end{subfigure}\hfill\par\medskip%
\begin{subfigure}{.5\columnwidth}
\includegraphics[width=\columnwidth]{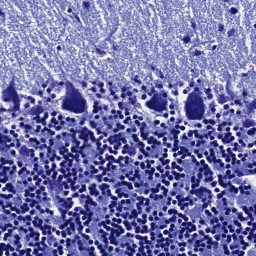}%
\caption{Cerebellum - Chimpanzee}%
\end{subfigure}\hfill%
\begin{subfigure}{.5\columnwidth}
\includegraphics[width=\columnwidth]{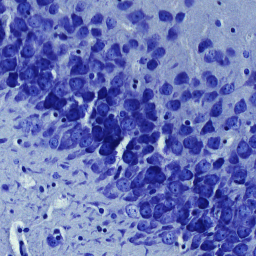}%
\caption{Hippocampus - Mouse}%
\end{subfigure}\hfill%
\begin{subfigure}{.5\columnwidth}
\includegraphics[width=\columnwidth]{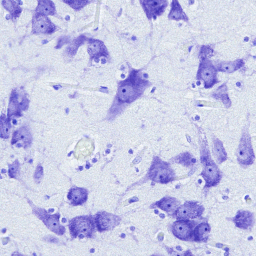}%
\caption{Auditory Cortex - B. dolphin}%
\label{dolphinimagecell}
\end{subfigure}\hfill%
\begin{subfigure}{.5\columnwidth}
\includegraphics[width=\columnwidth]{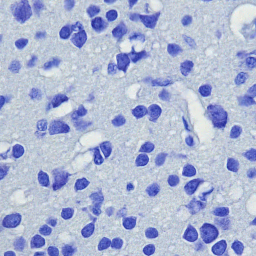}%
\caption{Visual Cortex - Macaque}%
\end{subfigure}%
\caption{Top: Morphological and density properties of cells in different brain areas from \textit{CytoDArk0}. Bottom: Sample images from \textit{CytoDArk0}. In particular, images are taken from \textit{CytoDArk0\_20x\_512}, a subset of \textit{CytoDArk0} with images at $20$x and size $512\times512$.}
\label{CytoDArkdesc}
\end{figure*}
\label{datasets}
The data used in this study belong to four datasets. 
\begin{itemize} 
\item \textit{PanNuke} \citep{gamper2020PanNuke} The largest pan-cancer nuclear instance segmentation and classification dataset, containing nearly $200,000$ labelled nuclei in  $7,901$ H\&E-stained histology images of size $256\times256$ from $19$ different tissue types at $40$x objective magnification. For each image, an instance segmentation and a classification mask is provided, where each nucleus is assigned to one of $5$ classes: Neoplastic, Epithelial, Connective/soft tissue cells, Inflammatory, Dead cells. 
\item \textit{CoNIC Challenge Dataset} \citep{graham2021lizard, graham2024CoNIC} This is the largest single tissue nuclear instance segmentation and classification dataset, containing nearly $500,000$ labelled nuclei in  $4,981$ H\&E-stained colon tissue histology images of size $256\times256$ at $20$x objective magnification from $5$ different data sources. Each nucleus is assigned to one of $6$ classes: Neutrophil, Epithelial, Lymphocyte, Plasma, Eosinophil, Connective tissue cells.

\item \textit{MoNuSeg} \citep{kumar2019multi} is an instance segmentation dataset containing nearly $7,000$ labelled nuclei in H\&E-stained histology images of size $1,000 \times 1,000$ from $7$ different tissue types  at several disease states (benign and tumors
at different stages)  at $40$x objective magnification. For each image, an instance segmentation mask is provided. In this work, we consider MoNuSeg as an external test set.
\item \textit{CytoDArk0} This paper introduces a novel dataset comprising $38,755$ labeled brain cells (neurons and glia) in Nissl-stained histology images. As detailed in Table \ref{tab:CytoDArksummary}, the dataset comprises a set of $69$ $1024\times1024$ RGB images at $20$x magnification (\textit{CytoDArk0\_20x\_1024}) sourced from four distinct brain regions (cerebellum, hippocampus, auditory cortex, visual cortex) across five different species (Bottlenose Dolphin, Mouse, Chimpanzee, Macaque, Bovine). Among these, a subset of $58$ images is also available at $40$x magnification, with size $2048\times2048$ (\textit{CytoDArk0\_40x\_2048}). We also share a patched version of the image sets, which we call \textit{CytoDArk0\_20x\_512} ($276$ patches), \textit{CytoDArk0\_20x\_256} ($1,104$ patches), \textit{CytoDArk0\_40x\_1024} ($232$ patches), \textit{CytoDArk0\_40x\_512} ($928$ patches), and \textit{CytoDArk0\_40x\_256} ($3,712$ patches). 
This dataset, named \textit{CytoDArk0}, is the first publicly available resource for cell instance segmentation in Nissl-stained histological brain images. It was created by processing brain samples from the University of Padova, collected in compliance with legal and ethical standards. The Nissl staining technique was used to consistently stain the entire population of cells in nervous tissue \citep{pilati2008rapid}. This technique specifically targets the rough endoplasmic reticulum and ribosomes within cell bodies, producing blue or purple cell bodies against a light or white background, as shown in Fig. \ref{CytoDArkdesc}. It is widely recognized as the most effective technique for assessing cell body morphology, density, and distribution in the brain, as well as for highlighting the regional or laminar organization of cytoarchitecture across different brain areas \citep{garcia2016distinction}. In contrast, H\&E staining, although commonly used in histopathology, employs double coloration that stains the nuclei of all cells blue or dark purple and other structures, such as the cytoplasm and extracellular matrix, in shades of pink. This approach reduces the contrast between cell borders and the extracellular matrix, and may not provide as clear or specific a visualization of brain cell populations. To offer an overview of the diverse cell populations in each distinct brain region, we extracted morphological and density information for each cell.
Fig. \ref{CytoDArkdesc} (top) shows the distributions of cell area, axis ratio (minor axis / major axis), solidity (area / convex hull area), and density (average number of cells within a radius of $55 \mu m$) for the different brain regions (CE: \textit{Cerebellum}, HI: \textit{Hippocampus}, AC: \textit{Auditory Cortex}, VC: \textit{Visual Cortex}). We limited the plots to data within the 99th percentile to reduce the effect of long tails.
Cells in all regions generally exhibit a regular shape (median solidity value higher than or equal to $0.95$) because Nissl staining highlights cell bodies rather than neuronal processes, which typically make neuron shapes appear less compact. The cerebellum data shows higher axis ratio and density values, and lower area values, consistent with the cerebellum high density of small, circular granule cells in the granular layer and larger, pear-shaped Purkinje cells in the Purkinje layer. The hippocampus data display a bimodal cell area distribution, with the second-highest cell density, reflecting its cytoarchitecture of densely packed granule cells in areas like the dentate gyrus and relatively larger pyramidal neurons. The cerebral cortex is usually characterized by a diverse range of neurons (such as pyramidal cells) and glial cells (including astrocytes, oligodendrocytes, and microglia), each with unique shapes and functions. In CytoDArk0, the smaller median cell area compared to other regions is mainly due to the presence of small, circular glial cells (visible in Fig. \ref{dolphinimagecell}). However, as shown in Fig. \ref{subfiga}, there is also a noticeable long tail, representing the diverse brain cell population. It is important to emphasize that the statistics presented do not represent entire brain areas, as they are based on sample patches from WSIs. Additionally, these statistics do not account for the distinct cytoarchitectural characteristics of different brain subregions, such as the layers in the cerebral cortex or cerebellum, which are not differentiated here. This contrasts with detailed cytoarchitectural studies, such as \citet{corain2020multi}. Nonetheless, these statistics still provide insights into some overall similarities and differences between brain areas. All images were annotated at the highest available resolution using QuPath \citep{bankhead2017qupath} to delineate the contours of neurons and glia. Initial annotations were made by the fourth and first author after proper training, followed by verification and corrections from our team of expert neuroanatomists. Such initial annotations were used in an active learning fashion to train CISCA predecessors—MR-NOM \citep{vadori2023mr} and NCIS\citep{vadori2023ncis}—and create preliminary annotations on additional images, subsequently reviewed and corrected by our team of experts, and used to fine-tune the models. Annotations were exported from QuPath and processed with Matlab and Python to generate the instance segmentation mask along with supplementary maps necessary to apply the methodology presented in this paper or similar ones, such as the contour mask and the distance masks. 
Most of the images are from the auditory cortex of the bottlenose dolphin. This is because initial research efforts were concentrated on this species and brain region. While fewer images are available for other brain regions, as discussed in  Sections \ref{results} and \ref{conclusions}, good detection and segmentation performances are achievable in these areas. This outcome is expected, given that the data from the auditory cortex include a broad range of brain cell sizes and shapes, and the images from the cerebellum and hippocampus contain a higher cell density, providing a greater number of cell examples for learning. By disseminating this first dataset version, we aim to facilitate progress in digital neuropathology and studies of brain cytoarchitecture. Future versions of CytoDArk may include an increased number of annotated images and the annotation of cell types, distinguishing between neurons and glia. CytoDArk0 is available at \url{https://doi.org/10.5281/zenodo.13694738}.
\end{itemize}
\subsection{Data Splits}
As reported in Table \ref{tab:split_composition}, each dataset used for training was divided into three subsets: ca. 70\% for training, ca. 10\% for validation, and ca. 20\% for testing. For Conic and PanNuke, these splits deviate from those described in the CoNIC Challenge \citep{graham2024CoNIC} or PanNuke dataset \citep{gamper2020PanNuke} because 1. the test set for the CoNIC Challenge is not openly available, which forced us to isolate a test set from the publicly available training data 2. PanNuke is officially split into $3$ folds for cross-validation purposes, which we are not applying in this paper, where we opt for a single split for each dataset. Details on replicating these splits will be available at \url{https://github.com/Vadori/CytoArk}. For MoNuSeg, the official test set was used as an external test set.
\section{Results}
\label{results}
\subsection{Evaluation Metrics}

To evaluate performance, we selected some mainstream measures and measures recently applied for the CoNIC challenge \citep{graham2024CoNIC} and PanNuke reference papers \citep{gamper2020PanNuke}, aiming to provide a comprehensive and unified perspective.

\begin{itemize}
\item \textit{Cell instance segmentation} performance can be assessed in three main aspects: distinguishing cell bodies from the background, detecting individual cell instances, and segmenting individual cell instances. 

For evaluating how well a model \textit{differentiates cell bodies from the background}, we use the \textit{Dice Coefficient} (\textit{\textbf{Dice}}) and the \textit{Average Jaccard Index} (\textit{\textbf{AJI}}). 
Both metrics evaluate the overlap between GT and predicted instances.  Dice considers predicted instances as a single foreground instance and measures the overlap with the GT as follows:
\begin{equation}
\textit{Dice} = \frac{2 \sum_{i,j=1}^{N} \mathbb{I}(\mathbf{Y}^{l}_{ij}) \odot \mathbb{I}(\hat{\mathbf{Y}}^{l}_{i,j})}{\sum_{i,j=1}^{N} \mathbb{I}(\mathbf{Y}^{l}_{ij})+\sum_{i,j=1}^{N}  \mathbb{I}(\hat{\mathbf{Y}}^{l}_{ij})}
\end{equation}
Here,  $\hat{\mathbf{Y}^{l}} \in \mathbb{N}^{N\times N}$ is the predicted label matrix and
the indicator function $\mathbb{I}$ is used to binarize the label matrices.
The AJI \citep{kumar2017dataset} extends the Jaccard Index by matching each GT instance with the predicted instance that yields the highest intersection over union (IoU). It sums the intersections of these matched pairs in the numerator and their union in the denominator. Additionally, the area of unmatched predicted instances is included in the denominator to account for false positives. Unlike Dice, AJI accounts for the precise matching between GT and predicted cells. As a result, AJI tends to be more stringent than Dice.

To assess the models ability to \textit{detect individual cell instances}, we consider the \textit{precision} (\textit{\textbf{P}}), \textit{recall} (\textit{\textbf{R}}), and $\textit{\textbf{F1}}$ \textit{score}. A predicted instance is compared to the GT, and a match (\textit{true positive}) is established if both centers of mass
are within a radius of $6$ pixels ($0.50 \mu m/$pixel) and $12$ pixels $(0.25 \mu m/$pixel),  for magnifications $20$x and $40$x, respectively \citep{sirinukunwattana2016locality}. 
Predicted instances that do not match are counted as false positives, while unmatched GT instances are considered false negatives. By indicating as $TP$, $FP$ and $FN$ the set of true positives, false positives and false negatives, respectively, the precision, recall and detection quality are  given by $P= |TP|/(|TP|+|FP|)$, $R=|TP|/(|TP|+|FN|)$, $F1=|TP|/(|TP|+0.5\cdot(|FP|+|FN|))$.

A similar approach is used to assess the models' ability to \textit{segment individual cell instances}. In this case, a predicted instance is compared to the GT, and a match is established if the IoU is grater than $0.5$. 
In line with \citet{graham2019hover}, we use the \textit{panoptic quality} (\textit{\textbf{PQ}}) defined as:
 \begin{equation}
 PQ =\frac{|TP|}{|TP|+0.5\cdot(|FP|+|FN|)}\cdot\frac{\sum_{(\hat{s}, s)\in TP} IoU(\hat{s}, s)}{|TP|}
 \label{eq:panoptic}
\end{equation}
where $IoU(\hat{s}, s)$ denotes the IoU between a matched predicted ($\hat{s}$) and GT ($s$) instance. The panoptic quality is comprised of two factors known as \textit{detection quality} ($DQ$) and \textit{segmentation quality} ($SQ$). DQ assesses the model's ability to locate cells accurately (i.e., with high area overlap with the GT), while SQ evaluates the model's capability to precisely outline the boundaries of these localized cells. 

All the above metrics are computed separately for each image, and the results are averaged.
\item \textit{Cell classification} performance can be evaluated with the \textit{multi-class panoptic quality} \textit{\textbf{mPQ+}} as defined in the CoNIC
challenge. A panoptic quality \textit{\textbf{PQ+}} is computed for each class separately by summing the TP, FP, and FN \textit{across all images}, applying the same formula as in Eq. \ref{eq:panoptic}. This method ensures that the metric is not affected by the absence of a cell class in some images. In the PanNuke reference paper, the multi-class panoptic quality \textit{\textbf{mPQ}} is defined differently: it is evaluated independently for each image and class, considering only classes present according to the GT. The results are then averaged across images for each distinct tissue type and subsequently averaged across all tissue types, ensuring that each tissue type is weighted equally. Consequently, for PanNuke, we report the $mPQ$ value in addition to the $mPQ+$.
For PanNuke and CytoDArk0, which encompass various tissue types and brain regions, respectively, we also display the \textit{binary PQ} (\textit{\textbf{bPQ}}), which is computed in the same way as $mPQ$ but assumes that all nuclei belong to one class. It’s important to note that $bPQ$ differs from $PQ$, as averages are computed first across tissue types (or brain regions).
\item \textit{Cell counting} performance represents the ability to determine \textit{cellular composition} \citep{graham2024CoNIC}. It can be quantified via the \textit{coefficient of determination} \textit{\textbf{R$^2$}} to assess the correlation between the predicted and true counts as:
 \begin{equation}
 R^2 = 1-\frac{RSS}{TSS}
 \label{eq:coeffdetermination}
\end{equation}
where $RSS$ stands for the sum of squares of residuals and $TSS$ stands for the total sum of squares after fitting a regression line to the predicted and actual counts. When class information is available, we compute R$^2$ for each class and then average the results. When classes are not defined, no average is taken. 
\end{itemize}
To provide  insights into the results on different cell types, we display a subset of the above metrics for each cell class, namely P, R, F1, PQ+, R$^2$, thus combining the metrics applied in the CoNIC challenge (PQ+, R$^2$) and PanNuke (P, R, F1).

\definecolor{custompurple}{rgb}{0.55, 0.0, 1}
\definecolor{customteal}{HTML}{0.C0C1}
\definecolor{lightgrey}{HTML}{D0D0D0}

\setulcolor{lightgrey}
\setul{0.7pt}{.4pt}

\fboxsep1pt
\begin{table*}[]
\captionsetup{aboveskip=4pt, belowskip=2pt}
\caption{Overall performance of instance segmentation and classification using the proposed CISCA method and SOTA approaches. All metric values are expressed as percentages, except for \#P, which is the count of network trainable parameters in millions. \textit{Summary} results indicate the percentage of datasets in which the model ranks either first or second for each metric. The best values are in bold, while the second-best values are underlined. \textit{*Baseline values from the CoNIC Challenge are reported for reference. Note that these values were computed on a test set that is not publicly available and are based on a different data split for training and validation.}}
\label{tab:performance}
\centering
\setlength{\tabcolsep}{5pt}  
\renewcommand{\arraystretch}{0.9}  
\begin{tabular}{@{}lllllllllllll@{}}
\toprule
 &  & \#P (mln) & Dice & AJI & P & R & F1 & DQ & SQ & PQ & mPQ+ & (m)R$^2$ \\ \midrule
\textbf{CoNIC} (20x) &  & \color[HTML]{656565}- & \color[HTML]{656565}- & \color[HTML]{656565}- & \color[HTML]{656565}- & \color[HTML]{656565}- & \color[HTML]{656565}- & \color[HTML]{656565}- & \color[HTML]{656565}- & \color[HTML]{656565}- & \color[HTML]{656565}37.2* & \color[HTML]{656565}48.4* \\
StarDist  & \color[HTML]{FFFFFF} xxxxxxxxx  & \ul{22.8} & \ul{80.3} & \ul{63.3} & 85.1 & \ul{78.2} & \textbf{81.5} & \textbf{81.3} & \ul{80.3} & \textbf{65.4} & \textbf{50.1} & \textbf{81.6} \\
Hover-Net  &  & 54.8 & 74.0 & 51.5 & \textbf{87.7} & 59.0 & 70.5 & 64.7 & 78.9 & 51.4 & \ul{37.3}  & 46.7 \\
CellViT  &  & 46.7 & 72.3 & 49.0 & \ul{87.0} & 56.0 & 68.2 & 61.2 & 78.9 & 48.9 & 32.9 & 35.2 \\					
CISCA &  & \textbf{22.5} & \textbf{82.7} & \textbf{64.9} & 82.2 & \textbf{80.0} & \ul{81.1} & \ul{80.6} & \textbf{80.7} & \ul{65.2}  & \textbf{50.1} & \ul{80.9}  \\ \midrule
\textbf{PanNuke} (40x) &  &  &  &  &  &  &  &  &  &  &  &  \\
StarDist  &  & \ul{22.8} & 80.0 & 63.6 & \textbf{84.6} & 75.7 & 79.9 & \textbf{77.9} & 80.6 & \ul{63.8} & 43.3 & 71.9 \\
Hover-Net  &  & 54.7 & 81.3 & 64.0 & 81.0 & 76.3 & 78.6 & 74.9 & 80.4 & 61.2 & 38.3 & 56.8 \\
CellViT  &  & 46.7 & \ul{82.2} & \ul{65.2} & 82.1 & \textbf{78.2} & \ul{80.1} & 77.1 & \ul{81.1} & 63.6 & \textbf{49.9} & \textbf{81.5} \\				
CISCA &  & \textbf{22.5} & \textbf{83.1} & \textbf{65.8} & \ul{82.6} & 78.1 & \textbf{80.3} & \ul{77.5} & \textbf{81.5} & \textbf{63.9} & \ul{45.9} & \ul{77.7} \\ \midrule
\textbf{CytoDArk0\_20x\_256} &  &  &  &  &  &  &  &  &  & &  &  \\
StarDist  &  & \ul{22.5} & 79.5 & 61.8 & 77.7 & 77.4 & 77.5 & 76.6 & 82.4 & 63.1 & - & \ul{98.9} \\
Hover-Net  &  & 45.0 & \ul{86.6} & 68.6 & \textbf{85.9} & 69.8 & 77.0 & 76.4 & \ul{85.7} & 65.7 & - & 87.3 \\
CPN  &  & 31.6 & 86.5 & \ul{72.8} & 77.5 & \textbf{93.4} & \textbf{84.7} & \ul{80.2} & 85.4 & \ul{68.5} & - & 94.2 \\					
CISCA &  & \textbf{22.3} & \textbf{88.9} & \textbf{75.6} & \ul{81.7} & \ul{84.9} & \ul{83.3} & \textbf{82.2} & \textbf{87.0} & \textbf{71.5} & - & \textbf{99.1} \\ \midrule
\textbf{CytoDArk0\_40x\_256} &  &  &  &  &  &  &  &  & &  &  &  \\
StarDist  &  & \ul{22.5} & 77.8 & 60.3 & 80.3 & 74.4 & 77.2 & 73.5 & 82.0 & 60.4 & - & 97.6 \\
Hover-Net  &  & 45.0 & \ul{86.3} & \ul{71.3} & \ul{82.0} & 81.8 & 81.9 & \ul{78.1} & \ul{85.6} & \ul{67.4} & - & \textbf{98.4} \\
CPN  &  & 31.6 & 83.9 & 69.4 & 72.5 & \textbf{94.5} & \ul{82.1} & 75.9 & 84.6 & 64.4 & - & 83.5 \\
CISCA &  & \textbf{22.3} & \textbf{87.6} & \textbf{74.0} & \textbf{86.4} & \ul{82.2} & \textbf{84.3} & \textbf{81.3} & \textbf{87.2} & \textbf{71.0} & - & \ul{98.3} \\ \arrayrulecolor{customteal}\midrule 
\textbf{Summary} (see caption) &  &  &  &  &  &  &  &  & &  &  &  \\ 
StarDist    &  & \textbf{100} & 25.0 & 25.0 & 25.0 & 25.0 & 25.0 & 50.0 & 25.0 & 50.0 & 50.0 & 50.0 \\
Hover-Net &  & 0 & 50.0 &25.0 & \textbf{75.0} & 0 & 0 & 25.0 & 50.0 & 25.0 &50.0 &25.0 \\
Cell-ViT    &  & 0 & 50.0 & 50.0 & 50.0 & 50.0 & 50.0 & 0 & 50.0 & 0 & 50.0 & 50.0 \\
CPN         &  & 0 & 0 &50.0 & 0 & \textbf{100} & \textbf{100} & 50.0 & 0 & 50.0 & - & 0 \\
CISCA      &  & \textbf{100} & \textbf{100} & \textbf{100} & \textbf{75.0} & \textbf{100} & \textbf{100} & \textbf{100} & \textbf{100} & \textbf{100} & \textbf{100} & \textbf{100} \\ \arrayrulecolor{customteal}\bottomrule
\end{tabular}

\end{table*}

\begin{table*}[]
\captionsetup{aboveskip=4pt, belowskip=2pt}
\caption{Performance on PanNuke by tissue type. \textit{*Baseline values from PanNuke are reported for reference. Note that these values were computed with a $3-$fold validation approach, differently from the train-validation-test split used in this study.}}
\label{tab:PanNuke_performance}
\centering
\setlength{\tabcolsep}{5pt}  
\renewcommand{\arraystretch}{0.9}  
\begin{tabular}{@{}llllllllllllllll@{}}
\toprule
 \multicolumn{16}{c}{\textbf{PanNuke} (40x)} \\ \midrule
 &  & \multicolumn{2}{c}{\color[HTML]{656565} Baseline*} & \multicolumn{1}{c}{} & \multicolumn{2}{c}{StarDist} & \multicolumn{1}{c}{} & \multicolumn{2}{c}{Hover-Net} & \multicolumn{1}{c}{} & \multicolumn{2}{c}{CellViT} & \multicolumn{1}{c}{} & \multicolumn{2}{c}{CISCA} \\ \cmidrule(lr){3-4} \cmidrule(lr){6-7} \cmidrule(lr){9-10} \cmidrule(lr){12-13} \cmidrule(l){15-16} 
 &  & \multicolumn{1}{c}{{\color[HTML]{656565} bPQ}} & \multicolumn{1}{c}{{\color[HTML]{656565} mPQ}} & \multicolumn{1}{c}{} & \multicolumn{1}{c}{bPQ} & \multicolumn{1}{c}{mPQ} & \multicolumn{1}{c}{} & \multicolumn{1}{c}{bPQ} & \multicolumn{1}{c}{mPQ} & \multicolumn{1}{c}{} & \multicolumn{1}{c}{bPQ} & \multicolumn{1}{c}{mPQ} & \multicolumn{1}{c}{} & \multicolumn{1}{c}{bPQ} & \multicolumn{1}{c}{mPQ} \\
Adrenal Gland &  & {\color[HTML]{656565} 69.6} & {\color[HTML]{656565}  48.1} &  & 67.5 & 46.0 &  & 66.9 & 40.7 &  & \textbf{69.5} & \textbf{48.9} &  & \ul{68.8} & \ul{46.2} \\
Bile Duct &  & {\color[HTML]{656565} 67.0} & {\color[HTML]{656565} 47.1} &  & \ul{68.5} & \ul{48.0} &  & 66.6 & 46.6 &  & 67.6 & \textbf{48.1} &  & \textbf{69.2} & 46.5 \\
Bladder &  & {\color[HTML]{656565} 70.3} & {\color[HTML]{656565} 57.9} &  & 70.3 & 50.7 &  & 69.7 & 52.9 &  & \ul{70.9} & \ul{53.7} &  & \textbf{72.2} & \textbf{54.0} \\
Breast &  & {\color[HTML]{656565} 64.7} & {\color[HTML]{656565} 49.0} &  & \ul{64.9} & \ul{44.3} &  & 62.1 & 43.1 &  & \textbf{65.1} & \textbf{48.7} &  & 64.4 & \ul{44.3} \\
Cervix &  & {\color[HTML]{656565} 66.5} & {\color[HTML]{656565} 44.4} &  & 66.3 & \ul{45.3} &  & 63.9 & 40.4 &  & \ul{66.6} & \textbf{45.9} &  & \textbf{68.8} & 43.3 \\
Colon &  & {\color[HTML]{656565} 55.8} & {\color[HTML]{656565} 41.0} &  & \textbf{56.8} & \ul{40.4} &  & 53.0 & 38.3 &  & 55.9 & \textbf{42.9} &  & \ul{56.7} & 39.8 \\
Esophagus &  & {\color[HTML]{656565} 64.3} & {\color[HTML]{656565} 50.9} &  & \ul{65.0} & \ul{47.9} &  & 61.7 & 44.7 &  & 64.2 & \textbf{51.0} &  & \textbf{65.3} & \textbf{51.0} \\
Head \&   Neck &  & {\color[HTML]{656565} 63.3} & {\color[HTML]{656565} 45.3} &  & \ul{60.0} & \textbf{40.1} &  & 58.1 & 39.2 &  & 58.4 & 38.4 &  & \textbf{60.6} & \ul{39.5} \\
Kidney &  & {\color[HTML]{656565} 68.4} & {\color[HTML]{656565} 44.2} &  & 70.5 & 39.8 &  & \ul{71.5} & \ul{44.6} &  & \textbf{72.0} & \textbf{50.9} &  & \textbf{72.0} & 43.3 \\
Liver &  & {\color[HTML]{656565} 72.5} & {\color[HTML]{656565} 49.7} &  & 70.1 & 48.4 &  & 68.8 & 47.1 &  & \ul{71.2} & \textbf{49.6} &  & \textbf{72.3} & \ul{49.2} \\
Lung &  & {\color[HTML]{656565} 63.0} & {\color[HTML]{656565} 40.0} &  & \textbf{61.5} & 37.0 &  & 58.1 & 36.1 &  & 59.9 & \textbf{37.4} &  & \ul{60.4} & \ul{37.1} \\
Ovarian &  & {\color[HTML]{656565} 63.1} & {\color[HTML]{656565} 48.6} &  & \textbf{64.2} & \textbf{47.7} &  & 60.3 & 44.1 &  & \ul{62.2} & \ul{47.5} &  & 61.7 & 44.4 \\
Pancreatic &  & {\color[HTML]{656565} 64.9} & {\color[HTML]{656565} 46.0} &  & \ul{64.6} & \textbf{47.9} &  & 62.1 & 41.6 &  & 63.3 & \ul{46.9} &  & \textbf{64.9} & 44.0 \\
Prostate &  & {\color[HTML]{656565} 66.2} & {\color[HTML]{656565} 51.0} &  & \textbf{64.9} & \textbf{47.1} &  & \ul{63.9} & 45.4 &  & 62.3 & \ul{46.5} &  & \ul{63.9} & \textbf{47.1} \\
Skin &  & {\color[HTML]{656565} 62.3} & {\color[HTML]{656565} 34.3} &  & \textbf{61.6} & \ul{36.7} &  & 58.2 & 33.4 &  & \ul{61.3} & \textbf{37.0} &  & 59.8 & 34.4 \\
Stomach &  & {\color[HTML]{656565} 68.9} & {\color[HTML]{656565} 47.3} &  & \ul{66.6} & \textbf{46.9} &  & 65.1 & 41.8 &  & 65.2 & 43.8 &  &  \textbf{67.6} & \ul{45.8} \\
Testis &  & {\color[HTML]{656565} 68.9} & {\color[HTML]{656565} 47.5} &  & \ul{66.2} & \ul{45.2} &  & 63.5 & 40.6 &  & 65.9 & \textbf{45.6} &  & \textbf{67.6} & 43.0 \\
Thyroid &  & {\color[HTML]{656565} 69.8} & {\color[HTML]{656565} 43.2} &  & 64.4 & 40.3 &  & 63.4 & \ul{41.6} &  & \textbf{67.0} & \textbf{42.4} &  & \ul{65.6} & 41.3 \\
Uterus &  & {\color[HTML]{656565} 63.9} & {\color[HTML]{656565} 43.9} &  & \textbf{62.4} & 37.4 &  & 56.5 & 37.7 &  & 60.1 & \ul{38.2} &  & \ul{60.7} & \textbf{42.3} \\ \midrule
Average across   tissues &  & {\color[HTML]{656565} 66.0} & {\color[HTML]{656565} 46.3} &  & \ul{65.1} & \ul{44.0} &  & 62.8 & 42.1 &  & 64.7 & \textbf{45.4} &  & \textbf{65.4} & \ul{44.0} \\
STD across   tissues &  & {\color[HTML]{656565} 3.8} & {\color[HTML]{656565} 5.0} &  & \textbf{3.6} & \textbf{4.4} &  & 4.8 & \textbf{4.4} &  & \ul{4.5} & 4.9 &  & 4.6 & \ul{4.7} \\ \bottomrule
\end{tabular}

\end{table*}
\begin{table*}[]
\captionsetup{aboveskip=4pt, belowskip=2pt}
\caption{Detection performance (top), panoptic quality and counting performance (bottom) for each cell type in CoNIC.}
\label{tab:conic_class_perf}

\centering
\setlength{\tabcolsep}{2pt}  
\renewcommand{\arraystretch}{1}  
\begin{tabular}{@{}lllllllllllllllllllllllll@{}}
\toprule
\multicolumn{25}{c}{\textbf{CoNIC} (20x)} \\ \midrule
 &  & \multicolumn{3}{c}{Neutrophil} & \multicolumn{1}{c}{} & \multicolumn{3}{c}{Epithelial} & \multicolumn{1}{c}{} & \multicolumn{3}{c}{Lymphocyte} & \multicolumn{1}{c}{} & \multicolumn{3}{c}{Plasma} & \multicolumn{1}{c}{} & \multicolumn{3}{c}{Eosinophil} & \multicolumn{1}{c}{} & \multicolumn{3}{c}{Connective} \\ \cmidrule(lr){3-5} \cmidrule(lr){7-9} \cmidrule(lr){11-13} \cmidrule(lr){15-17} \cmidrule(lr){19-21} \cmidrule(l){23-25} 
 &  & \multicolumn{1}{c}{P} & \multicolumn{1}{c}{R} & \multicolumn{1}{c}{F1} & \multicolumn{1}{c}{} & \multicolumn{1}{c}{P} & \multicolumn{1}{c}{R} & \multicolumn{1}{c}{F1} & \multicolumn{1}{c}{} & \multicolumn{1}{c}{P} & \multicolumn{1}{c}{R} & \multicolumn{1}{c}{F1} & \multicolumn{1}{c}{} & \multicolumn{1}{c}{P} & \multicolumn{1}{c}{R} & \multicolumn{1}{c}{F1} & \multicolumn{1}{c}{} & \multicolumn{1}{c}{P} & \multicolumn{1}{c}{R} & \multicolumn{1}{c}{F1} & \multicolumn{1}{c}{} & \multicolumn{1}{c}{P} & \multicolumn{1}{c}{R} & \multicolumn{1}{c}{F1} \\
StarDist &  & 29.9 & \textbf{37.2} & \ul{33.1} &  & 79.8 & \ul{74.1} & \textbf{76.8} &  & \textbf{67.8} & \textbf{65.1} & \textbf{66.4} &  & 48.3 & \textbf{34.2} & \textbf{40.1} &  & 44.1 & \textbf{39.8} & \textbf{41.9} &  & \textbf{66.6} & \ul{62.1} & \textbf{64.3} \\
Hover-Net & {\color[HTML]{FFFFFF} xx} & 34.6 & 19.4 & 24.9 & {\color[HTML]{FFFFFF} xx} & \textbf{82.2} & 63.3 & 71.5 & {\color[HTML]{FFFFFF} xx} & 65.1 & 32.7 & 43.5 & {\color[HTML]{FFFFFF} xx} & \ul{51.2} & 22.7 & 31.4 & {\color[HTML]{FFFFFF} xx} & \textbf{52.2} & 24.5 & 33.4 & {\color[HTML]{FFFFFF} xx} & \ul{65.4} & 42.9 & 51.8 \\
CellViT &  & \textbf{40.4} & 8.2 & 13.6 &  & \ul{81.5} & 64.0 & 71.7 &  & 65.5 & 29.4 & 40.6 &  & 44.8 & 20.9 & 28.5 &  & \ul{51.0} & 16.7 & 25.1 &  & 61.3 & 36.9 & 46.1 \\
 CISCA &  & \ul{36.6} & \ul{35.0} & \textbf{35.8} &  & 76.5 & \textbf{75.4} & \ul{75.9} &  & \ul{67.3} & \ul{63.6} & \ul{65.4} &  & \textbf{53.8} & \ul{30.7} & \ul{39.1} &  & 42.2 & \ul{39.1} & \ul{40.6} &  & 60.1 & \textbf{67.1} & \ul{63.4} \\ \midrule
 &  & \multicolumn{1}{c}{PQ+} & \multicolumn{1}{c}{} & \multicolumn{1}{c}{R2} & \multicolumn{1}{c}{} & \multicolumn{1}{c}{PQ+} & \multicolumn{1}{c}{} & \multicolumn{1}{c}{R2} & \multicolumn{1}{c}{} & \multicolumn{1}{c}{PQ+} & \multicolumn{1}{c}{} & \multicolumn{1}{c}{R2} & \multicolumn{1}{c}{} & \multicolumn{1}{c}{PQ+} & \multicolumn{1}{c}{} & \multicolumn{1}{c}{R2} & \multicolumn{1}{c}{} & \multicolumn{1}{c}{PQ+} & \multicolumn{1}{c}{} & \multicolumn{1}{c}{R2} & \multicolumn{1}{c}{} & \multicolumn{1}{c}{PQ+} & \multicolumn{1}{c}{} & \multicolumn{1}{c}{R2} \\
{\color[HTML]{656565} Baseline} & {\color[HTML]{656565} } & {\color[HTML]{656565} 13.4} & {\color[HTML]{656565} } & {\color[HTML]{656565} 27.7} & {\color[HTML]{656565} } & {\color[HTML]{656565} 49.2} & {\color[HTML]{656565} } & {\color[HTML]{656565} 56.1} & {\color[HTML]{656565} } & {\color[HTML]{656565} 50.9} & {\color[HTML]{656565} } & {\color[HTML]{656565} 52.9} & {\color[HTML]{656565} } & {\color[HTML]{656565} 38.7} & {\color[HTML]{656565} } & {\color[HTML]{656565} 44.2} & {\color[HTML]{656565} } & {\color[HTML]{656565} 29.4} & {\color[HTML]{656565} } & {\color[HTML]{656565} 57.8} & {\color[HTML]{656565} } & {\color[HTML]{656565} 41.8} & {\color[HTML]{656565} } & {\color[HTML]{656565} 51.8} \\
StarDist &  & \ul{30.7} &  & \ul{65.0} &  & \textbf{62.0} &  & \ul{95.6} &  & \ul{65.9} &  & \textbf{97.0} &  & \textbf{47.5} &  & \textbf{74.4} &  & \ul{36.4} &  & \textbf{68.4} &  & \textbf{57.8} &  & \ul{89.1} \\
Hover-Net &  & 23.1 &  & 47.7 &  & 51.3 &  & 74.4 &  & 41.4 &  & 33.9 &  & 34.7 &  & 33.3 &  & 28.1 &  & 38.9 &  & 45.0 &  & 52.1 \\
CellViT &  & 13.5 &  & 17.6 &  & 50.7 &  & 77.3 &  & 39.4 &  & 38.1 &  & 32.4 &  & 25.8 &  & 22.0 &  & 18.9 &  & 39.6 &  & 33.8 \\
CISCA &  & \textbf{32.1} &  & \textbf{69.5} &  & \ul{61.1} &  &\textbf{96.4} &  & \textbf{66.3} &  & \ul{96.2} &  & \ul{47.0} &  & \ul{68.1} &  & \textbf{36.5} &  & \ul{66.1} &  & \ul{57.3} &  & \textbf{89.4} \\ \bottomrule
\end{tabular}

\end{table*}

\begin{table*}[]
\captionsetup{aboveskip=4pt, belowskip=2pt}
\caption{Detection performance (top), panoptic quality and counting performance (bottom) for each cell type in PanNuke.}
\label{tab:PanNuke_class_perf}
\captionsetup{aboveskip=12pt, belowskip=4pt}
\centering
\setlength{\tabcolsep}{3pt}  
\renewcommand{\arraystretch}{1}  
\begin{tabular}{@{}lllllllllllllllllllll@{}}
\toprule
 \multicolumn{21}{c}{\textbf{PanNuke} (40x)} \\ \midrule
 &  & \multicolumn{3}{c}{Neoplastic} & \multicolumn{1}{c}{} & \multicolumn{3}{c}{Inflammatory} & \multicolumn{1}{c}{} & \multicolumn{3}{c}{Connective} & \multicolumn{1}{c}{} & \multicolumn{3}{c}{Dead} & \multicolumn{1}{c}{} & \multicolumn{3}{c}{Non-neop./Epith.} \\ \cmidrule(lr){3-5} \cmidrule(lr){7-9} \cmidrule(lr){11-13} \cmidrule(lr){15-17} \cmidrule(l){19-21} 
 &  & \multicolumn{1}{c}{P} & \multicolumn{1}{c}{R} & \multicolumn{1}{c}{F1} & \multicolumn{1}{c}{} & \multicolumn{1}{c}{P} & \multicolumn{1}{c}{R} & \multicolumn{1}{c}{F1} & \multicolumn{1}{c}{} & \multicolumn{1}{c}{P} & \multicolumn{1}{c}{R} & \multicolumn{1}{c}{F1} & \multicolumn{1}{c}{} & \multicolumn{1}{c}{P} & \multicolumn{1}{c}{R} & \multicolumn{1}{c}{F1} & \multicolumn{1}{c}{} & \multicolumn{1}{c}{P} & \multicolumn{1}{c}{R} & \multicolumn{1}{c}{F1} \\
{\color[HTML]{656565} Baseline} & {\color[HTML]{656565} } & {\color[HTML]{656565} 60.5} & {\color[HTML]{656565} 49.5} & {\color[HTML]{656565} 54.4} & {\color[HTML]{656565} } & {\color[HTML]{656565} 47.9} & {\color[HTML]{656565} 38.0} & {\color[HTML]{656565} 42.4} & {\color[HTML]{656565} } & {\color[HTML]{656565} 45.1} & {\color[HTML]{656565} 38.3} & {\color[HTML]{656565} 41.5} & {\color[HTML]{656565} } & {\color[HTML]{656565} 34.6} & {\color[HTML]{656565} 8.2} & {\color[HTML]{656565} 13.3} & {\color[HTML]{656565} } & {\color[HTML]{656565} 35.1} & {\color[HTML]{656565} 49.8} & {\color[HTML]{656565} 41.2} \\
StarDist &  & 61.1 & 50.7 & 55.4 &  & 49.4 & \ul{50.0} & \ul{49.7} &  & \ul{47.9} & 37.9 & 42.3 &  & 29.0 & \ul{35.1} & 31.8 &  & 39.9 & 48.7 & 43.9 \\
Hover-Net & \color[HTML]{FFFFFF} xxxx & 60.5 & 49.5 & 54.4 &  & 47.9 & 38.0 & 42.4 &  & 45.1 & 38.3 & 41.5 &  & \ul{34.6} & 8.2 & 13.3 &  & 35.1 & 49.8 & 41.2 \\
CellViT &  & \textbf{66.4} & \textbf{65.8} & \textbf{66.1} &  & \ul{59.4} & \textbf{50.3} & \textbf{54.5} &  & \textbf{51.8} & \ul{47.3} & \textbf{49.5} &  & \textbf{50.0} & 27.8 & \textbf{35.8} &  & \textbf{65.6} & \textbf{69.3} & \textbf{67.4} \\
CISCA &  & \ul{64.7}& \ul{55.8} & \ul{60.0} &  & \textbf{60.7} & 39.1 & 47.5 &  & 39.7 & \textbf{55.8} & \ul{46.4} &  & 34.2 & \textbf{35.5} & \ul{34.8} &  & \ul{63.2} & \ul{51.3} & \ul{56.6} \\ \midrule
 &  & \multicolumn{1}{c}{PQ+} & \multicolumn{1}{c}{} & \multicolumn{1}{c}{R2} & \multicolumn{1}{c}{{\color[HTML]{FFFFFF} xx}} & \multicolumn{1}{c}{PQ+} & \multicolumn{1}{c}{} & \multicolumn{1}{c}{R2} & \multicolumn{1}{c}{{\color[HTML]{FFFFFF} xx}} & \multicolumn{1}{c}{PQ+} & \multicolumn{1}{c}{} & \multicolumn{1}{c}{R2} & \multicolumn{1}{c}{{\color[HTML]{FFFFFF} xx}} & \multicolumn{1}{c}{PQ+} & \multicolumn{1}{c}{} & \multicolumn{1}{c}{R2} & \multicolumn{1}{c}{{\color[HTML]{FFFFFF} xx}} & \multicolumn{1}{c}{PQ+} & \multicolumn{1}{c}{} & \multicolumn{1}{c}{R2} \\
StarDist &  & 53.9 &  & 75.9 &  & \ul{51.0} &  & \textbf{90.5} &  & 40.9 &  & 55.4 &  & 26.3 &  & \textbf{78.5} &  & 44.3 &  & 59.4 \\
Hover-Net &  & 52.1 &  & 74.7 &  & 45.5 &  & 70.0 &  & 39.4 &  & 52.8 &  & 13.2 &  & 32.9 &  & 41.4 &  & 53.7 \\
CellViT &  & \textbf{59.4} &  & \textbf{86.9} &  & \textbf{55.3} &  & \textbf{90.5} &  & \textbf{45.2} &  & \textbf{73.8} &  & \textbf{31.0} &  & 70.2 &  & \textbf{58.6} &  & \textbf{85.9} \\
CISCA &  & \ul{56.3} &  & \ul{83.8} &  & 50.2 &  & \ul{82.4} &  & \ul{43.4} &  & \ul{63.3} &  & \ul{27.9} &  & \ul{76.4} &  & \ul{51.8} &  & \ul{82.7} \\ \bottomrule
\end{tabular}

\end{table*}

\begin{table*}[]
\captionsetup{aboveskip=4pt, belowskip=2pt}
\caption{Detection and panoptic quality performance on CytoDArk0\_20x\_256.}
\label{tab:CytoDArk20_class_perf}
\centering
\setlength{\tabcolsep}{4pt}  
\renewcommand{\arraystretch}{1}  
\begin{tabular}{@{}llllllllllllllllllll@{}}
\toprule
  \multicolumn{20}{c}{\textbf{CytoDArk0\_20x\_256}} \\ \midrule
 & \multicolumn{4}{c}{StarDist} & \multicolumn{1}{c}{} & \multicolumn{4}{c}{Hover-Net} & \multicolumn{1}{c}{} & \multicolumn{4}{c}{CPN} & \multicolumn{1}{c}{} & \multicolumn{4}{c}{CISCA} \\ \cmidrule(lr){2-5} \cmidrule(lr){7-10} \cmidrule(lr){12-15} \cmidrule(l){17-20} 
 & \multicolumn{1}{c}{P} & \multicolumn{1}{c}{R} & \multicolumn{1}{c}{F1} & \multicolumn{1}{c}{bPQ} & \multicolumn{1}{c}{} & \multicolumn{1}{c}{P} & \multicolumn{1}{c}{R} & \multicolumn{1}{c}{F1} & \multicolumn{1}{c}{bPQ} & \multicolumn{1}{c}{} & \multicolumn{1}{c}{P} & \multicolumn{1}{c}{R} & \multicolumn{1}{c}{F1} & \multicolumn{1}{c}{bPQ} & \multicolumn{1}{c}{} & \multicolumn{1}{c}{P} & \multicolumn{1}{c}{R} & \multicolumn{1}{c}{F1} & \multicolumn{1}{c}{bPQ} \\
Auditory   Cortex & 76.1 & 73.3 & 74.0 & 61.2 &  & \textbf{82.9} & 65.9 & 72.5 & 61.4 &  & 76.0 & \textbf{89.9} & \textbf{81.8} & \ul{66.3} &  & \ul{80.5} & \ul{83.5} & \ul{81.3} & \textbf{69.2} \\
Cerebellum & 78.4 & 78.9 & 78.6 & 68.1 &  & \textbf{93.5} & 65.0 & 76.5 & 55.2 &  & 81.0 & \textbf{95.4} & \textbf{87.5} & 70.6 &  & \ul{85.9} & \ul{84.7} & \ul{85.2} & \textbf{75.4} \\
Hippocampus & \textbf{85.9} & 77.1 & 80.7 & 68.3 &  & 85.2 & 74.9 & 79.3 & 70.8 &  & 84.1 & \textbf{92.8} & \textbf{88.2} & 76.8 &  & \ul{85.6} & \ul{84.3} & \ul{84.7} & \textbf{77.5} \\
Visual Cortex & 69.8 & 80.4 & 74.0 & 63.8 &  & \textbf{78.6} & 80.9 & 79.0 & 70.6 &  & 69.8 & \textbf{94.6} & 79.7 & 69.4 &  & \ul{75.0} & \ul{89.0} & \textbf{80.7} & \textbf{72.6} \\ \midrule
AVG across   tissues & 77.6 & 77.4 & 76.8 & 65.3 &  & \textbf{85.0} & 71.7 & 76.8 & 64.5 &  & 77.7 & \textbf{93.2} & \textbf{84.3} & 70.7 &  & \ul{81.7} & \ul{85.4} & \ul{83.0} & \textbf{73.7} \\
STD across   tissues & 6.7 & \ul{3.1} & 3.4 & \textbf{3.5} &  & \ul{6.2} & 7.6 & \ul{3.1} & 7.6 &  & \ul{6.2} & \textbf{2.4} & 4.2 & 4.4 &  & \textbf{5.1} & \textbf{2.4} & \textbf{2.3} & \ul{3.6} \\ \bottomrule
\end{tabular}

\end{table*}

\begin{table*}[]
\captionsetup{aboveskip=4pt, belowskip=2pt}
\caption{Detection and panoptic quality performance on CytoDArk0\_40x\_256.}
\label{tab:CytoDArk40_class_perf}
\centering
\setlength{\tabcolsep}{4pt}  
\renewcommand{\arraystretch}{1}  
\begin{tabular}{@{}llllllllllllllllllll@{}}
\toprule
 \multicolumn{20}{c}{\textbf{CytoDArk0\_40x\_256}} \\ \midrule
 & \multicolumn{4}{c}{StarDist} & \multicolumn{1}{c}{} & \multicolumn{4}{c}{Hover-Net} & \multicolumn{1}{c}{} & \multicolumn{4}{c}{CPN} & \multicolumn{1}{c}{} & \multicolumn{4}{c}{CISCA} \\ \cmidrule(lr){2-5} \cmidrule(lr){7-10} \cmidrule(lr){12-15} \cmidrule(l){17-20} 
 & \multicolumn{1}{c}{P} & \multicolumn{1}{c}{R} & \multicolumn{1}{c}{F1} & \multicolumn{1}{c}{bPQ} & \multicolumn{1}{c}{} & \multicolumn{1}{c}{P} & \multicolumn{1}{c}{R} & \multicolumn{1}{c}{F1} & \multicolumn{1}{c}{bPQ} & \multicolumn{1}{c}{} & \multicolumn{1}{c}{P} & \multicolumn{1}{c}{R} & \multicolumn{1}{c}{F1} & \multicolumn{1}{c}{bPQ} & \multicolumn{1}{c}{} & \multicolumn{1}{c}{P} & \multicolumn{1}{c}{R} & \multicolumn{1}{c}{F1} & \multicolumn{1}{c}{bPQ} \\
Auditory   Cortex & 77.0 & 72.1 & 72.7 & 59.4 &  & \ul{78.8} & 79.5 & 77.7 & \ul{66.5} &  & 71.4 & \textbf{92.3} & \ul{79.1} & 64.2 &  & \textbf{83.4} & \ul{81.0} & \textbf{80.8} & \textbf{69.9} \\
Cerebellum & 80.1 & 78.6 & 78.9 & 68.5 &  & \ul{87.2} & 84.2 & \ul{85.0} & \ul{74.6} &  & 68.2 & \textbf{96.4} & 79.1 & 66.3 &  & \textbf{89.7} & \ul{86.2} & \textbf{87.7} & \textbf{79.3} \\ \midrule
AVG across   tissues & 78.5 & 75.4 & 75.8 & 63.9 &  & \ul{83.0} & 81.9 & \ul{81.3} & \ul{70.6} &  & 69.8 & \textbf{94.4} & 79.1 & 65.3 &  & \textbf{86.6} & \ul{83.6} & \textbf{84.3} & \textbf{74.6} \\
STD across   tissues & \textbf{2.2} & 4.5 & \ul{4.4} & 6.4 &  & 6.0 & \ul{3.3} & 5.2 & \ul{5.7} &  & \ul{2.3} & \textbf{2.9}& \textbf{0.0} & \textbf{1.5} &  & 4.5 & 3.7 & 4.9 & 6.6 \\ \bottomrule
\end{tabular}

\end{table*}

\subsection{Quantitative Results}
\label{sec:experiments}
We conducted a comprehensive comparative analysis of CISCA-Net, StarDist, Hover-Net, CellViT, and CPN on the CoNIC, PanNuke, and CytoDArk0 datasets. 
\begin{itemize}
\item \textbf{StarDist} \citep{schmidt2018cell, weigert2022nuclei} utilizes a U-Net backbone with 3 heads for binary pixel classification, regression of $64$ radial distance maps, and cell type classification. The outputs are post-processed to create contour proposals, which are filtered via NMS and converted into a label map via rasterization. StarDist recently achieved first place in the CoNIC challenge in terms of mPQ+ \citep{graham2024CoNIC}. 
\item \textbf{Hover-Net} \citep{graham2019hover} employs a Preact-ResNet50 backbone, followed by 3 decoders for binary pixel classification, regression of horizontal and vertical distance maps, and cell type classification. The outputs are post-processed using morphological operations to produce the label map. Hover-Net was used to set the baseline performance for the CoNIC Challenge and the PanNuke dataset \citep{gamper2020PanNuke}.
\item \textbf{CellViT} \citep{horst2024cellvit} utilizes a U-Net-like architecture with a ViT encoder pre-trained on 104 million $256 \times 256$ histological image patches from The Cancer Genome Atlas. This is followed by 3 decoders for binary pixel classification, regression of horizontal and vertical distance maps, and cell type classification. Post-processing is the same as in Hover-Net. \citet{horst2024cellvit} demonstrated CellViT's cell instance segmentation and classification SOTA performance on the PanNuke dataset.
\item \textbf{CPN} \citep{upschulte2022contour} uses a U-Net backbone with 4 heads: one for binary pixel classification, and the others for the regression of 20 maps for contour representation, 2 maps for contour localization, and 2 maps for contour refinement. The outputs are post-processed to create contour proposals, which are filtered using NMS and converted into a label map through rasterization. \citet{upschulte2022contour} demonstrated CPN's cell instance segmentation SOTA performance considering, among others, a dataset of gray-scale images of neuronal cell bodies stained with a modified Merker stain.
\end{itemize}
In all these approaches, identified instances are classified into cell types using majority voting.

Each model was trained once (or fine-tuned in the CellViT case) on each dataset. For CytoDArk0, we evaluated both 20x and 40x magnifications. During training, we used the CytoDArk0\_20x\_1024 and CytoDArk0\_40x\_1024 datasets and random cropping was performed on each $1024 \times 1024\times3$ image to produce a $256 \times 256\times3$ image for network input. For validation and testing, we used the CytoDArk0\_20x\_256 and CytoDArk0\_40x\_256 datasets, respectively. Note that CellViT was not applied to CytoDArk0 because its current implementation does not provide a straightforward parameterization to bypass cell type prediction. As CytoDArk0 lacks cell type information, modifying the model's architecture would have been necessary for training. Conversely, CPN  was not used on CoNIC and PanNuke due to its design being solely for instance segmentation. All models used for comparison were retrained under the same conditions for data loading and optimization. Specifically, open-source implementations provided by the original authors were modified to use the same oversampled training set, stain augmentation and training strategies as outlined in Sections \ref{sec:Oversampling}, \ref{sec:StainNormalizationAugmentation}, and \ref{sec:Training} while preserving the original model architecture, loss functions and post-processing approach. The only exception was made for CellViT, where the encoder was frozen for the first 25 epochs and unfrozen afterward. The same framework as the original authors was used: TensorFlow for Hover-Net and StarDist, and PyTorch for CellViT and CPN. No tissue type information was shared with the models, necessitating the removal of the tissue classification head from CellViT. 

The overall results are shown in Table \ref{tab:performance}, where the best values are in bold, while the second-best values are underlined. On the CoNIC (20x) dataset, CISCA achieved the highest Dice and AJI scores. Although CISCA's P is 6.3\% lower than Hover-Net's, it has the highest R. The discrepancy between P and R is the smallest among all methods, leading to the second-highest F1 score, slightly below StarDist's. CISCA stays competitive across DQ, SQ, and PQ metrics, nearly matching StarDist, while Hover-Net and CellViT lag behind. The mPQ+ score is the same as StarDist's, indicating SOTA performance for multi-class cell instance segmentation. For cell counting, CISCA's mR$^2$ score is less than 1\% lower than StarDist's. The CoNIC Challenge baseline values are reported for reference. Note that they are based on a undisclosed test set and a different data split.

On PanNuke (40x), CISCA again leads in Dice and AJI. StarDist excels in P, while CISCA and CellViT are closely aligned on R and F1. For DQ, SQ, and PQ, all models except Hover-Net are fairly aligned. CellViT leads in mPQ+ and mR$^2$. Generally, while CISCA and StarDist perform comparably on CoNIC and PanNuke, Hover-Net and CellViT show better performance on PanNuke. Baseline values computed with a $3-$fold validation approach are reported for reference.

On CytoDArk0\_20x\_256, CISCA outperforms other methods in Dice and AJI. Hover-Net scores highest in P, while CPN achieves the highest R. CISCA's F1 score is the second highest after CPN. Notably, the discrepancy between P and R is lower for CISCA compared to CPN. CISCA outperforms other methods in DQ, PQ, and R$^2$.
On CytoDArk0\_40x\_256, CISCA's performance is the best across all metrics, except for R and R$^2$, where CPN and Hover-Net excel, respectively.

In the same Table \ref{tab:performance}, the \textit{Summary} section reports the percentage of datasets in which a model ranks either first or second for each metric. For CellViT and CPN, only the 2 datasets on which the models were tested were considered. CISCA has the lowest parameter count and consistently ranks as the best or second-best method across all metrics in the datasets of this study.
\begin{table*}[]
\captionsetup{aboveskip=4pt, belowskip=2pt}
\caption{Performance of CISCA trained on PanNuke and applied to MoNuSeg as an external validation dataset.}
\label{tab:monuseg}
\centering
\begin{tabular}{@{}lllllllllll@{}}
\toprule
  \multicolumn{11}{c}{\textbf{MoNuSeg}} \\ \midrule
 & \multicolumn{1}{c}{\# Param} & \multicolumn{1}{c}{Dice} & \multicolumn{1}{c}{AJI} & \multicolumn{1}{c}{P} & \multicolumn{1}{c}{R} & \multicolumn{1}{c}{F1} & \multicolumn{1}{c}{DQ} & \multicolumn{1}{c}{SQ} & \multicolumn{1}{c}{PQ} & \multicolumn{1}{c}{R2} \\
CISCA (trained on PanNuke) & 22.5 & 81.0 & 63.5 & 83.4 & 87.9 & 85.6 & 84.7 & 74.1 & 62.8 & 93.1 \\ \bottomrule
\end{tabular}
\end{table*}

\begin{table*}[]
\captionsetup{aboveskip=4pt, belowskip=2pt}
\caption{Ablation analysis conducted on CytoDArk0\_20x\_256 to evaluate the impact of various components and configurations within CISCA-Net.}
\label{tab:ablation}
\centering
\begin{tabular}{@{}lllllllllll@{}}
\toprule
  \multicolumn{11}{c}{\textbf{Ablation Analysis}} \\ \midrule
 & \# Param & Dice & AJI & P & R & F1 & DQ & SQ & PQ & R2 \\
a) CISCA-Net w/o diag., 2 px   classes & \textbf{22.2} & \textbf{89.0} & 74.3 & 81.2 & 83.8 & 82.5 & 81.6 & 86.9 & 71.0 & 99.1 \\
b) CISCA-Net w/o diag., 3 px classes & \textbf{22.2} & 88.1 & 73.8 & 80.8 & 84.6 & 82.7 & 81.5 & 86.4 & 70.4 & 99.1 \\
c) CISCA-Net w/ diag., 2 px classes & \textbf{22.2} & 88.4 & 74.0 & 81.1 & 84.7 & 82.9 & 81.3 & 86.6 & 70.5 & 99.1 \\
d) CISCA-Net w/ diag., 3 px classes (proposed) & \textbf{22.2} & 88.9 & \textbf{75.6} & \textbf{81.7} & \textbf{84.9} & \textbf{83.3} & \textbf{82.2} & \textbf{87.0} & \textbf{71.5} & 99.1 \\
e) CISCA-Net w/ diag., 3 px classes + att. & 23.3 & 88.1 & 74.2 & 80.2 & \textbf{84.9} & 82.5 & 81.5 & 86.2 & 70.3 & 99.1 \\
f) CISCA-Net w/ diag., 3 px classes - augmentation & \textbf{22.2} & 88.1 & 73.6 & 80.9 & 83.4 & 82.1 & 79.8 & 86.8 & 69.3 & \textbf{99.2}  \\ \bottomrule
\end{tabular}
\end{table*}
\begin{figure}
\centering
\begin{subfigure}{\columnwidth}
\includegraphics[width=\columnwidth]{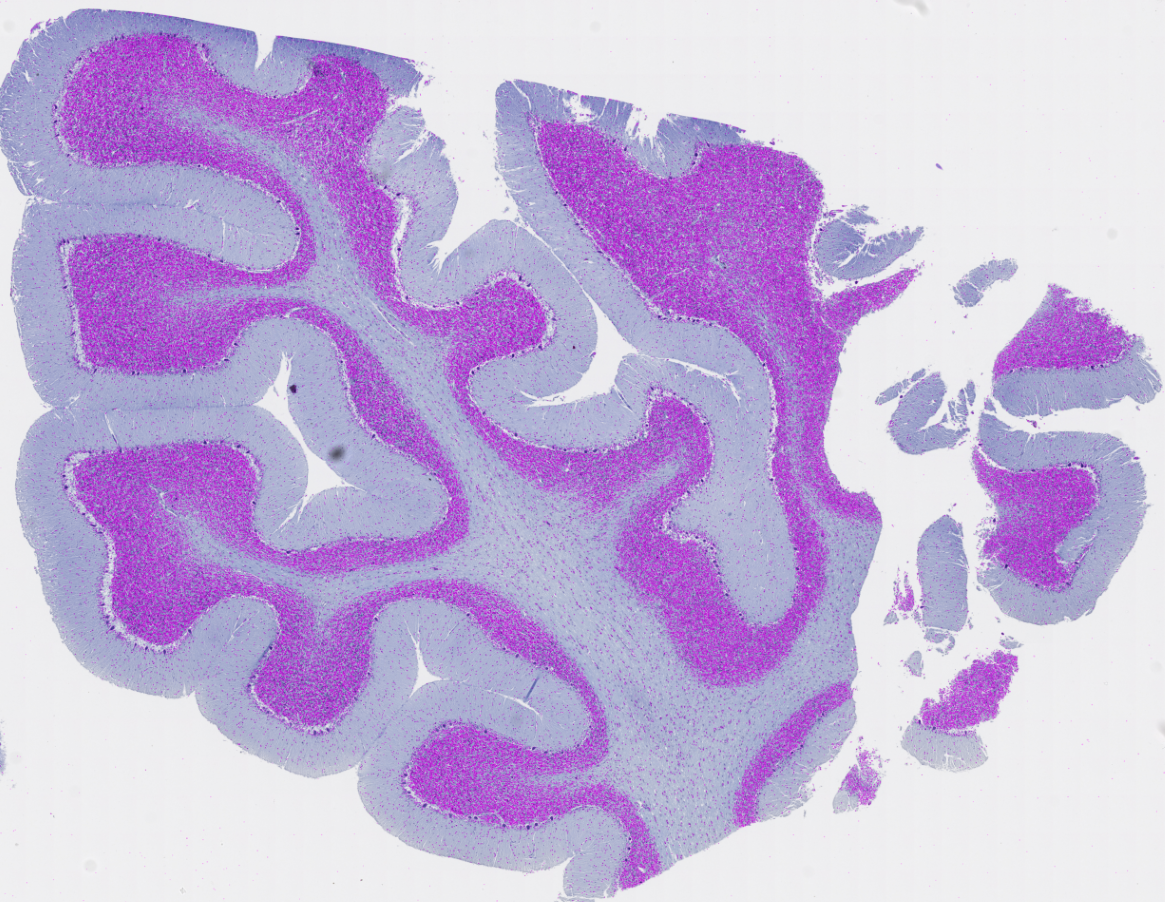}%
\end{subfigure}\hfill %
\begin{subfigure}{0.424\columnwidth}
\includegraphics[width=\columnwidth]{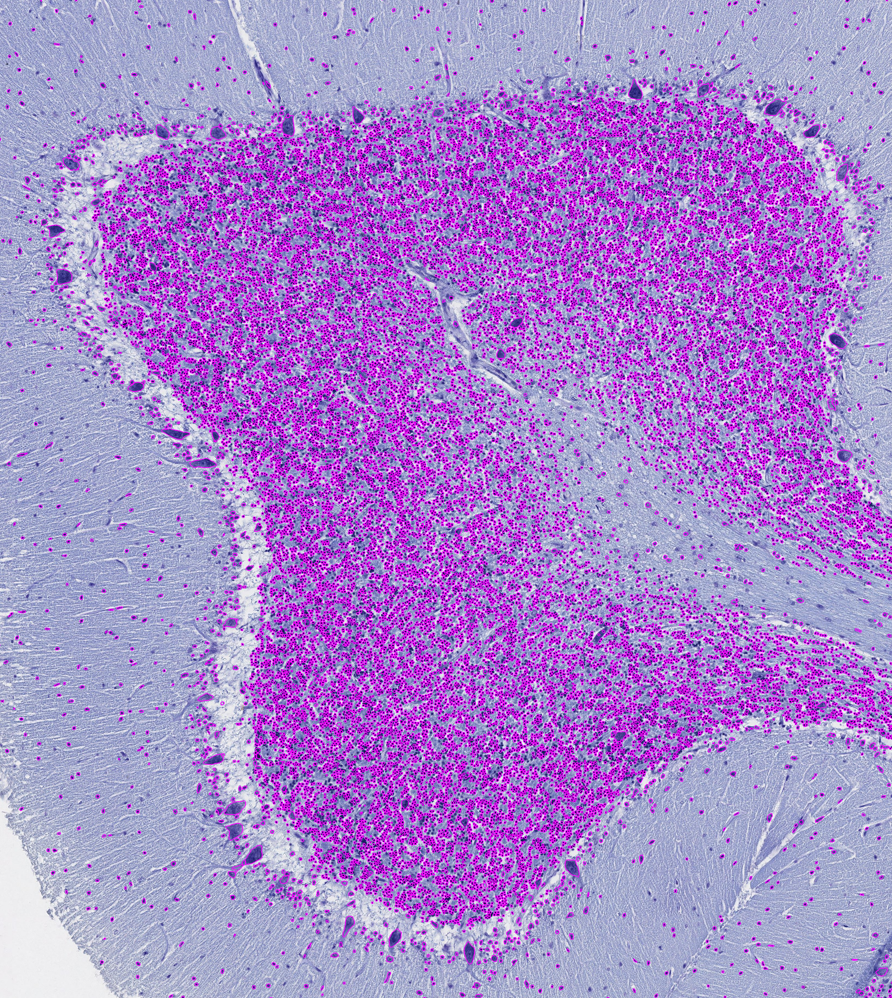}%
\end{subfigure}\hfill
\begin{subfigure}{0.571\columnwidth}
\includegraphics[width=\columnwidth]{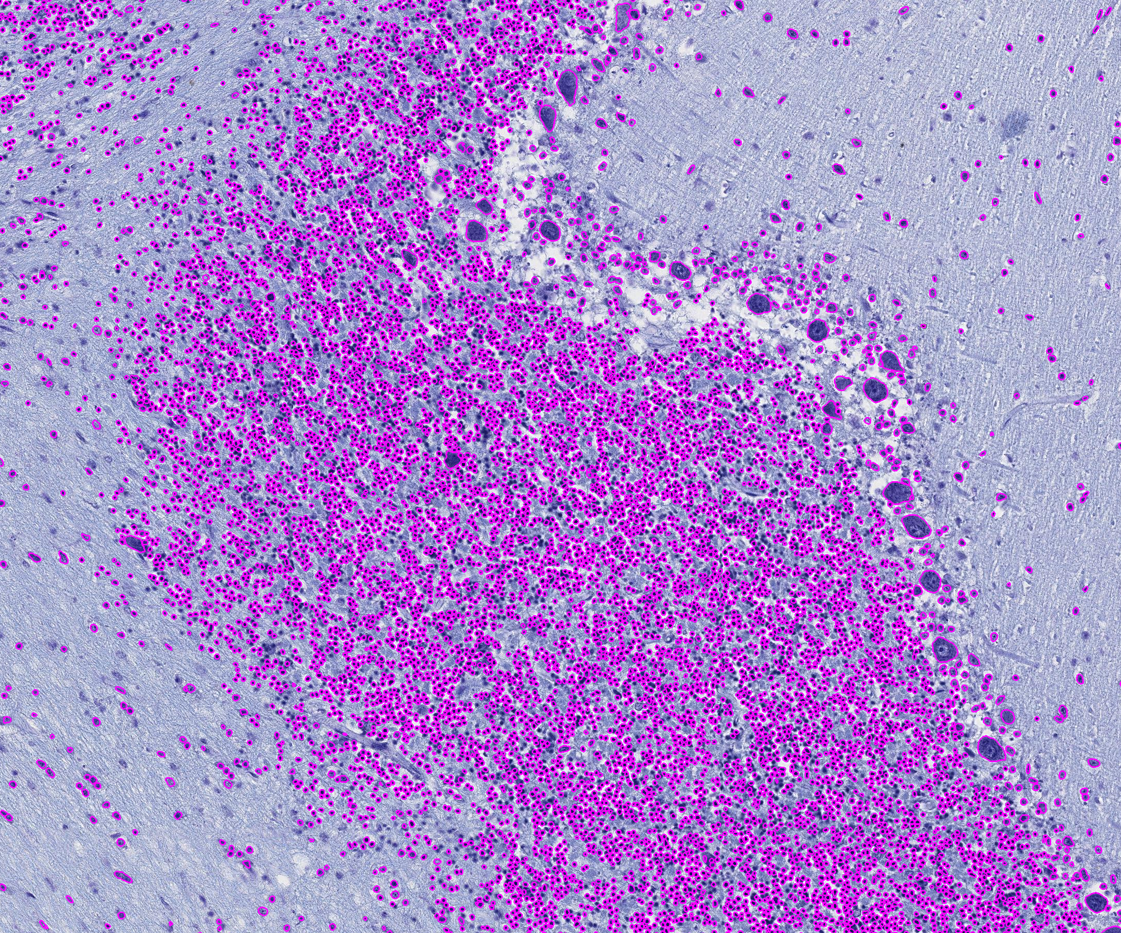}
\end{subfigure}
\caption{Qualitative results of CISCA applied to a WSI from the chimpanzee cerebellum, with details in the bottom row.}
\label{wsiexample}
\end{figure}

The results on PanNuke, detailed by tissue type, are shown in Table \ref{tab:PanNuke_performance}. Although performance varies across individual tissues, StarDist and CISCA achieve comparable overall results, both scoring the highest in bPQ. CellViT shows superior performance in terms of mPQ. These results align with the rankings observed in the previous PQ and mPQ+ metrics.

Table \ref{tab:conic_class_perf} shows the performance on CoNIC across different cell types. StarDist demonstrates superior detection performance, obtaining the highest F1 and R scores for most types, although its P consistently exceeds R, except for Neutrophils. Either StarDist or CISCA attain the highest PQ+ and R$^2$ values. In contrast, Hover-Net and CellViT demonstrate significantly lower recall for most cell types, leading to generally lower F1 scores. Additionally, Hover-Net and CellViT have lower PQ+ and R$^2$ scores compared to StarDist and CISCA.

Table \ref{tab:PanNuke_class_perf} shows the performance on PanNuke across cell types. Here, CellViT generally outperforms other models. CISCA ranks second in F1, PQ+, and R$^2$ for 4 out of 5 cell types.

Table \ref{tab:CytoDArk20_class_perf} shows the performance on CytoDArk0\_20x\_256 across different brain areas. CPN achieves the highest F1 on 3 out of 4 brain areas due to the low number of false negatives as documented by the very high R, while CISCA outperforms other methods in terms of bPQ across all brain areas and F1 on the visual cortex. Metrics tend to be generally lower on the auditory cortex, where there is significant variability among cells (cf. Section \ref{datasets}). It is noteworthy that although the models were not exposed to any images of the visual cortex during training, as these images were reserved for testing, their performance is comparable to or surpasses that achieved on the auditory cortex. 

Table \ref{tab:CytoDArk40_class_perf} presents the performance on CytoDArk0\_40x\_256 across different brain areas. Here, CISCA outperforms the other models across all metrics except R.

\subsection{External Validation}

We utilized the MoNuSeg dataset for external validation to assess the robustness of CISCA's performance. The same CISCA model trained on PanNuke was applied to MoNuSeg. The performance results are shown in Table \ref{tab:monuseg}. It can be observed that all metrics align with the results on PanNuke as shown in Table \ref{tab:performance}. Notably, F1 is higher (85.6\% vs. 80.3\%), DQ is higher (84.7\% vs. 77.5\%), SQ is lower (74.1\% vs. 81.5\%), and PQ is less than 2\% lower (62.8\% vs. 63.9\%).

\subsection{Ablation Study}

We conducted an ablation analysis on CytoDArk0\_20x\_256 to assess the effects of various CISCA-Net configurations, focusing on the following variants:
\begin{enumerate}[label=\alph*)]
\item \label{item:a} \textbf{CISCA-Net w/o diag. 2 px classes}: no diagonal distance map (only vertical and horizontal) and binary pixel classification (foreground vs. background), like Hover-Net.
\item \label{item:b} \textbf{CISCA-Net w/o diag. 3 px classes}: no diagonal and ternary pixel classification (BD vs. CB vs. BG).
\item \label{item:c} \textbf{CISCA-Net w/ diag. 2 px classes}: with diagonals and binary pixel classification.
\item \label{item:d} \textbf{CISCA-Net w/ diag. 3 px classes (proposed)}: with diagonals and ternary pixel classification.
\item \label{item:e} \textbf{CISCA-Net w/ diag. 3 px classes + att.}: with diagonals, ternary pixel classification, and attention gates in the skip connections, essentially substituting the U-Net backbone with Attention U-Net \citep{oktay2018attention}.
\item \label{item:f} \textbf{CISCA-Net w/ diag. 3 px classes - stain augmentation}: with diagonals and ternary prediction of pixel class without stain augmentation (cf. Section \ref{sec:StainNormalizationAugmentation}).
\end{enumerate}
Evaluation metrics for these models are reported in Table \ref{tab:ablation}. A comparison of \ref{item:a} with \ref{item:b} and \ref{item:c} with \ref{item:d} suggests that the boundary class is beneficial when diagonal distance maps are present. Comparing \ref{item:a} with \ref{item:c} and \ref{item:b} with \ref{item:d} shows that incorporating the diagonals enhances performance in the ternary class scenario. Therefore, the optimal performance is achieved with the combination of diagonals and ternary pixel classification, which is the proposed configuration. Switching to configurations \ref{item:e} with attention or \ref{item:f} without stain augmentation results in a decline in most metrics.

\subsection{Qualitative Results}
\label{sec:qualit}
In Fig. \ref{wsiexample}, we present qualitative results of CISCA applied to a WSI from the chimpanzee cerebellum. Neurons and glia cells are accurately identified and delineated, enabling precise cell counting and the extraction of morphological and density features from all cells in the WSI for comparative cytoarchitecture studies \citep{AMUNTS20071061, falcone2021neuronal,  graic2023cytoarchitectureal, graic2024age}.

\begin{figure*}
\centering
\begin{subfigure}{\textwidth}
\includegraphics[width=\columnwidth]{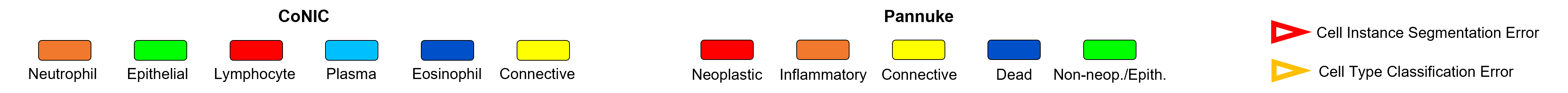}%
\end{subfigure}\hfill \hfil%
\begin{subfigure}{\textwidth}
\includegraphics[width=\columnwidth]{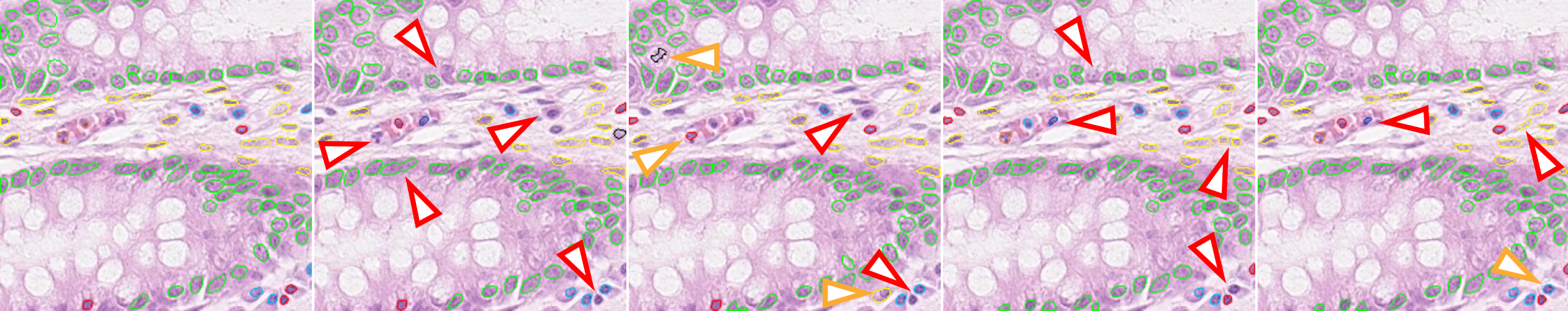}%
\end{subfigure}\hfill \hfil%
\begin{subfigure}{\textwidth}
\includegraphics[width=\columnwidth]{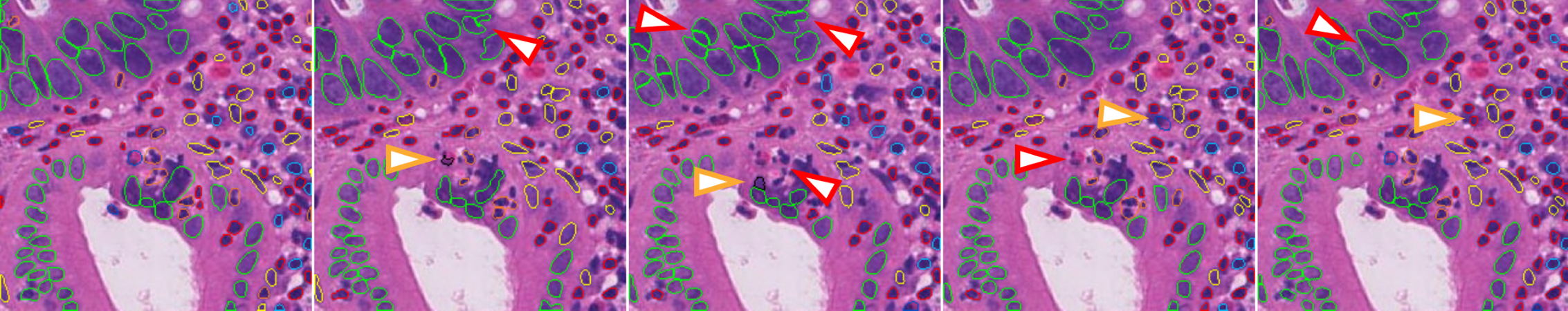}%
\end{subfigure}\hfill\hfil%
\begin{subfigure}{\textwidth}
\includegraphics[width=\columnwidth]{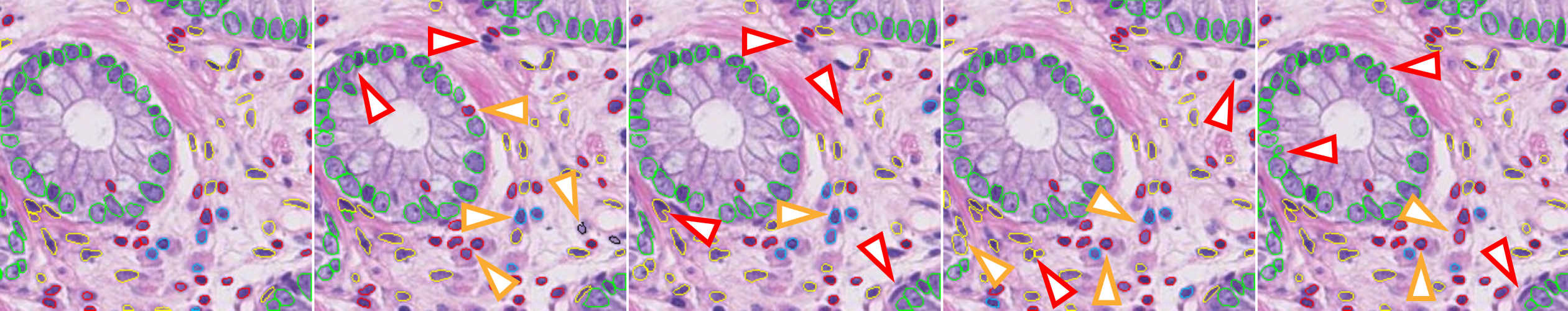}%
\end{subfigure}\hfill\hfil%
\vspace{0.3em}
\begin{subfigure}{\textwidth}
\includegraphics[width=\columnwidth]{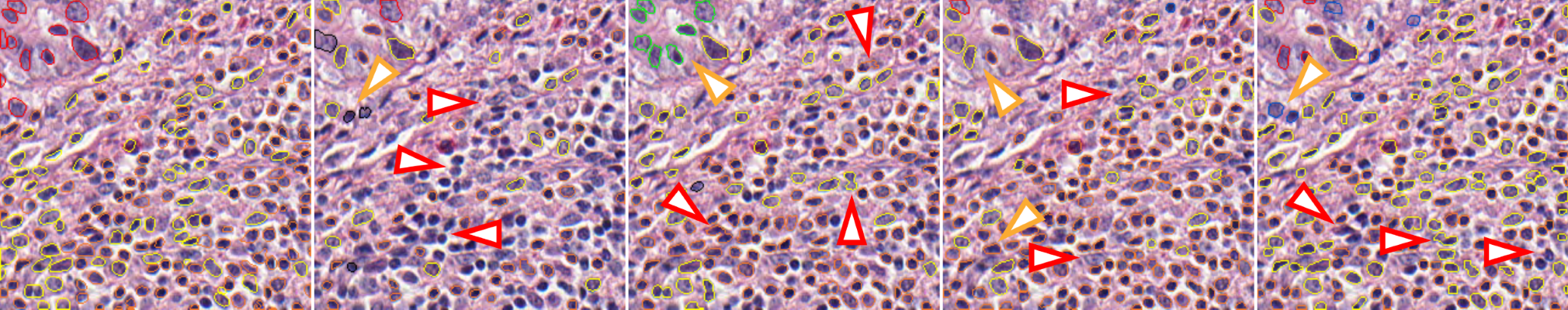}%
\end{subfigure}\hfill\hfil%
\begin{subfigure}{\textwidth}
\includegraphics[width=\columnwidth]{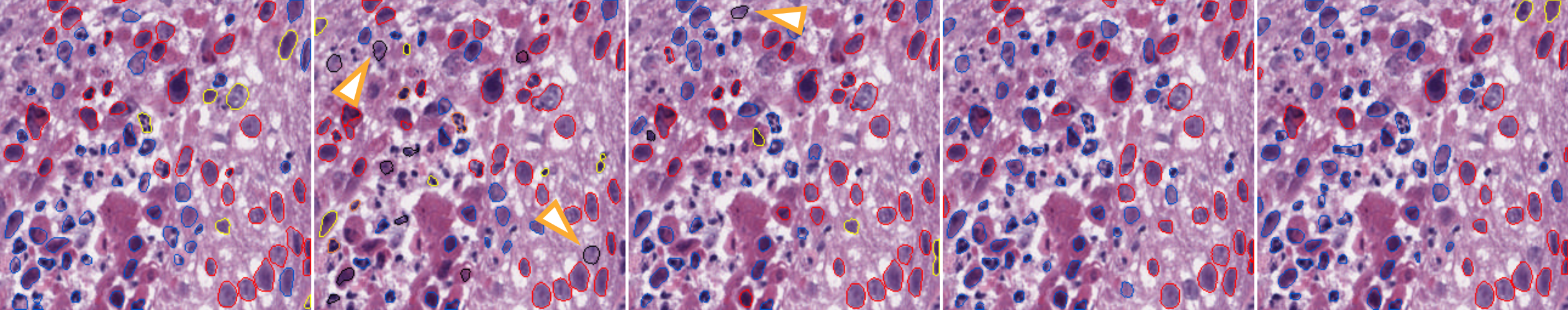}%
\end{subfigure}\hfill\hfil%
\begin{subfigure}{\textwidth}
\includegraphics[width=\columnwidth]{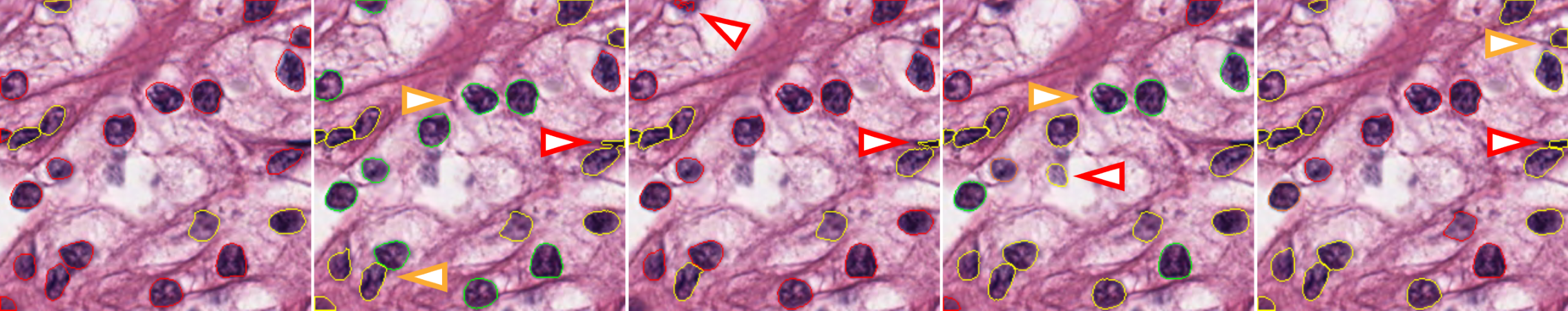}%
\end{subfigure}\hfill%
\par
\begin{subfigure}{.4\columnwidth}
\includegraphics[width=\columnwidth]{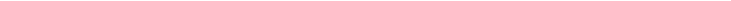}%
\caption{GT}%
\end{subfigure}\hfill%
\begin{subfigure}{.4\columnwidth}
\includegraphics[width=\columnwidth]{Figures/padline.png}%
\caption{Hover-Net}%
\end{subfigure}\hfill%
\begin{subfigure}{.4\columnwidth}
\includegraphics[width=\columnwidth]{Figures/padline.png}%
\caption{CellViT}%
\end{subfigure}\hfill%
\begin{subfigure}{.4\columnwidth}
\includegraphics[width=\columnwidth]{Figures/padline.png}%
\caption{StarDist}%
\end{subfigure}\hfill%
\begin{subfigure}{.4\columnwidth}
\includegraphics[width=\columnwidth]{Figures/padline.png}%
\caption{CISCA}%
\end{subfigure}\hfill%
\caption{Qualitative results for sample patches from CoNIC (top $3$ rows) and PanNuke (bottom $3$ rows). GT in the first column.}
\label{figabc}
\end{figure*}

\begin{figure*}
\centering
\begin{subfigure}{\textwidth}
\includegraphics[width=\columnwidth]{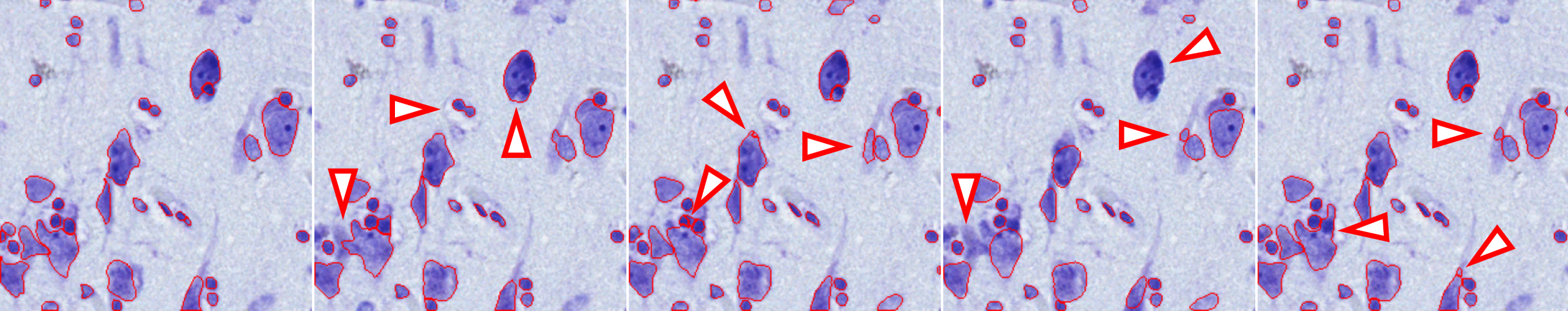}%
\end{subfigure}\hfill \hfil%
\begin{subfigure}{\textwidth}
\includegraphics[width=\columnwidth]{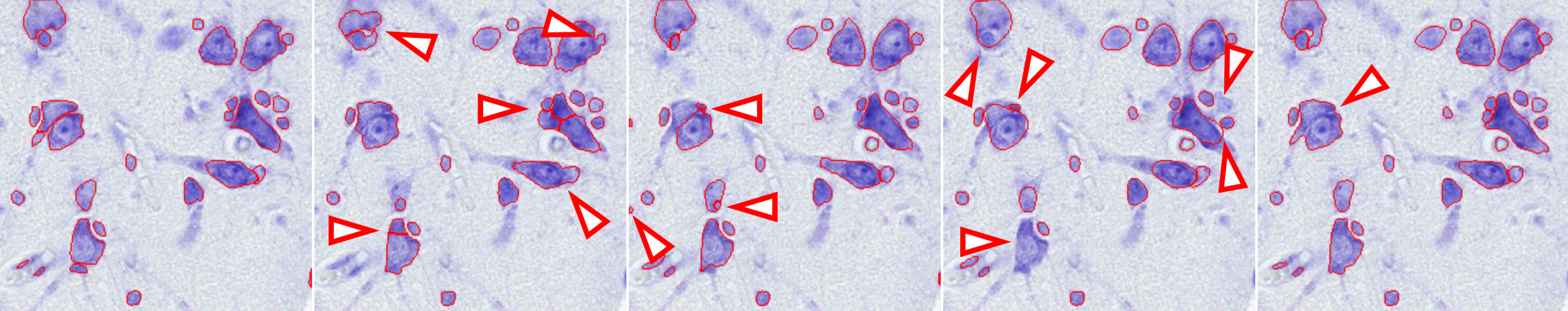}%
\end{subfigure}\hfill\hfil%
\begin{subfigure}{\textwidth}
\includegraphics[width=\columnwidth]{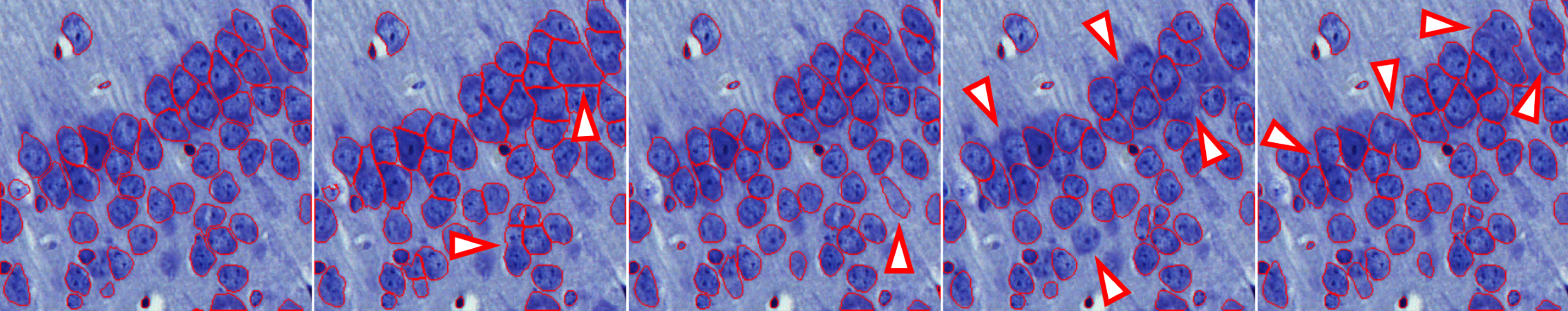}%
\end{subfigure}\hfill\hfil%
\vspace{0.3em}
\begin{subfigure}{\textwidth}
\includegraphics[width=\columnwidth]{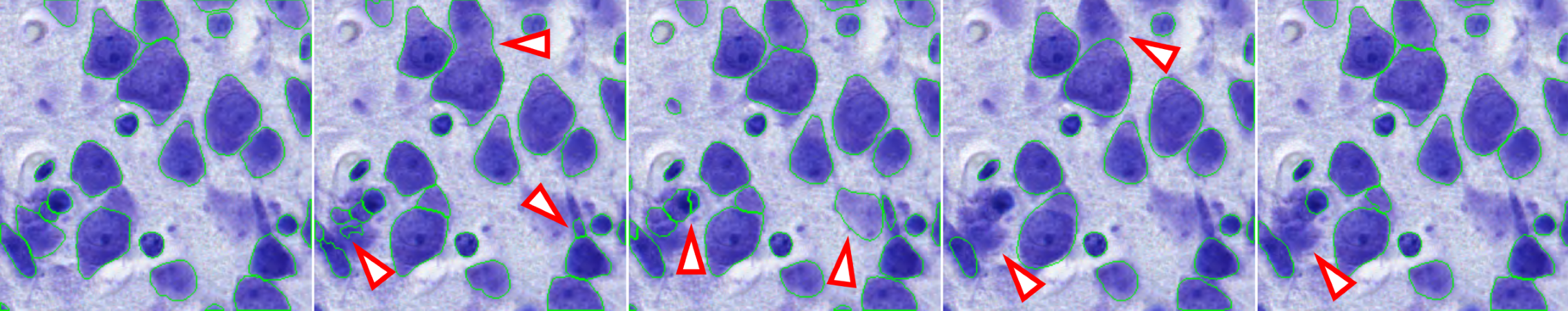}%
\end{subfigure}\hfill\hfil%
\begin{subfigure}{\textwidth}
\includegraphics[width=\columnwidth]{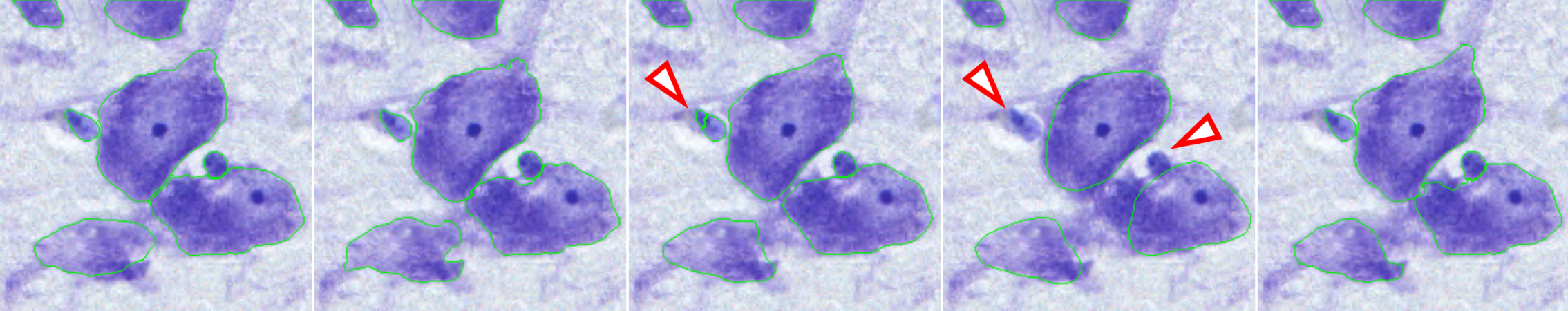}%
\end{subfigure}\hfill\hfil%
\begin{subfigure}{\textwidth}
\includegraphics[width=\columnwidth]{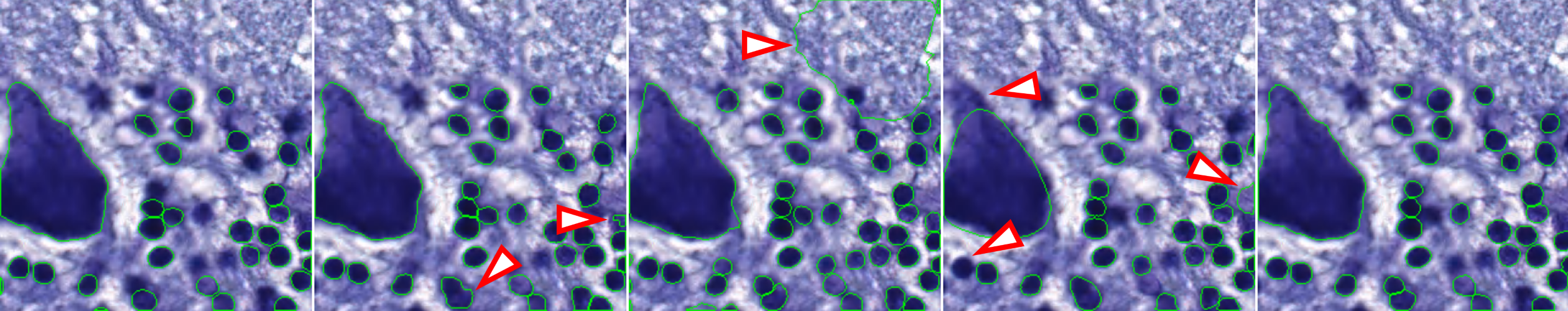}%
\end{subfigure}\hfill%
\par
\begin{subfigure}{.4\columnwidth}
\includegraphics[width=\columnwidth]{Figures/padline.png}%
\caption{GT}%
\end{subfigure}\hfill%
\begin{subfigure}{.4\columnwidth}
\includegraphics[width=\columnwidth]{Figures/padline.png}%
\caption{Hover-Net}%
\end{subfigure}\hfill%
\begin{subfigure}{.4\columnwidth}
\includegraphics[width=\columnwidth]{Figures/padline.png}%
\caption{CPN}%
\end{subfigure}\hfill%
\begin{subfigure}{.4\columnwidth}
\includegraphics[width=\columnwidth]{Figures/padline.png}%
\caption{StarDist}%
\end{subfigure}\hfill%
\begin{subfigure}{.4\columnwidth}
\includegraphics[width=\columnwidth]{Figures/padline.png}%
\caption{CISCA}%
\end{subfigure}\hfill%
\caption{Qualitative results for sample patches from CytoDArk0\_20x\_256 (top, red) and CytoDArk0\_40x\_256 (bottom, green). GT in the first column.}
\label{qualit:cyto}
\end{figure*}

Fig.  \ref{figabc} presents the qualitative results for CoNIC and PanNuke, while Fig.  \ref{qualit:cyto} shows the outcomes for CytoDArk0. Red arrows point out instance segmentation errors, whereas yellow arrows highlight cell type classification errors. CISCA's segmentations appear generally aligned with the GT. However, while CISCA typically separates clusters of cells effectively, there are instances—especially when the image is blurred—where CISCA fails to properly divide them. Other errors include over-segmentation, where smaller cells are erroneously created from noise or the tips of larger cells. 
StarDist performs well on PanNuke and CoNIC but is less accurate on CytoDArk, where its segmentations are overly smooth and do not accurately follow the actual cell contours. While StarDist tends to avoid cell fusion or over-segmentation, its primary errors involve completely missing some cells.
HoverNet and CellViT share similar visual characteristics due to CellViT’s reliance on HoverNet-like predictions and the same post-processing pipeline. However, HoverNet tends to introduce more unnatural cell shapes. Moreover, although all models exhibit some misclassification errors, CellViT and HoverNet seem to be the only ones that occasionally misclassify cells as background.
CPN’s contours are generally accurate and closely match GT. Notably, CPN performs exceptionally well on a sample of the hippocampus (Fig. \ref{qualit:cyto}, third row), where cells are tightly clustered.
In other samples, artifacts or specific textural patterns in the background are sometimes mistaken for cells.

\section{Discussion and Conclusions}
\label{conclusions}

In this study, we introduced CISCA, a novel pipeline designed for cell instance segmentation and classification in histo(patho)logical images, alongside CytoDArk0, the first open annotated dataset for cell instance segmentation in Nissl-stained 2D histological brain images. CISCA's core is CISCA-Net, a DL architecture with a lightweight U-Net backbone, followed by three convolutional heads. The first two heads handle instance segmentation by predicting four distance maps in four directions and classifying pixels into cell boundaries, cell bodies, and background. These outputs are processed through a custom pipeline, where the only user-defined hyperparameter is the threshold for the minimum cell instance size, denoted as $\theta_2$ (cf. Section \ref{sec:post}), which can be set to default values. The third head is dedicated to classifying the segmented cells. As demonstrated in Section \ref{results}, CISCA consistently outperforms other methods across multiple metrics on the datasets considered. As detailed in Table \ref{tab:performance}, CISCA ranks as the best or second-best method across all metrics in the four datasets analyzed in this study. It typically achieves the highest scores in Dice, AJI, PQ, and bPQ, reflecting its strong capability to distinguish foreground from background and accurately segment individual cells. It also maintains balanced and competitive P and R, contributing to high F1 scores and aligning with SOTA methods. Furthermore, CISCA excels in classification and counting tasks. Notably, CISCA demonstrates superior performance compared to HoverNet, despite their similarities. This advantage is partly due to the novel diagonal distance maps and ternary classification that CISCA introduces. The diagonal distance maps improve the detection of boundaries between closely positioned cells, especially when these boundaries are rotated relative to the vertical and horizontal axes. Moreover, the inclusion of a boundary class in the pixel classification task enables the first two heads to collaborate on boundary detection, fostering a synergistic mechanism that enhances prediction robustness.
Although CISCA-Net lacks the closed contour notion—which appears to be particularly beneficial for detecting tightly clustered cells that differ in appearance from those in the training set, such as those in hippocampal sections (cf. Section \ref{sec:qualit})—our methodology performs competitively with, or surpasses, StarDist and CPN, especially when considering panoptic quality measures. A significant portion of CISCA's success can be attributed to its post-processing pipeline, which is critical for effectively managing the model's output at both 20x and 40x resolutions. Compared with all the SOTA methods considered in this study, it features the lowest parameter count and avoids the computationally expensive NMS used by StarDist and CPN, making it an efficient solution.

StarDist excels in the digital pathology datasets but struggles with our Nissl-stained dataset due to the greater variance in cell sizes and shapes. The star-convex polygon proposal-based model tends to regularize cell shapes, making it less effective when cells deviate from mostly circular or oval forms, as seen in CoNIC or PanNuke.

Hover-Net is generally outperformed on digital pathology datasets but performs competitively on CytoDArk0 at 40x magnification. Meanwhile, CellViT, which utilizes a ViT pre-trained on millions of histology images in its encoder, excels on PanNuke at 40x magnification but underperforms on CoNIC at 20x. Additionally, both models sometimes classify cells as background (cf. Section \ref{sec:qualit}), potentially due to independent decoders for instance segmentation and classification that do not work in tandem as effectively as in other methods. Focusing on CellViT, the smaller dataset size of CoNIC, which targets a single tissue type (colon), may lead to overfitting. This is especially relevant since transformers require large datasets to fully utilize their flexible architecture. Typically, transformers enhance performance when trained on extremely large datasets, such as those in the computer vision field with hundreds of millions of images \citep{caron2021emerging,he2022masked}. When trained on `small' datasets of around 1 million images, the performance of vision transformers is comparable to that of CNNs \citep{dosovitskiy2020image}. However, as demonstrated in this study, training on `very small' datasets with a few thousand images can result in worse performance than CNNs, highlighting the Occam's Razor principle \citep{mackay1992bayesian}. In the biological context, collecting datasets of millions of diverse images may be unfeasible, suggesting that it may still be premature to claim that vision transformers surpass CNNs in cellular segmentation \citep{stringer2024transformers}, despite recent outstanding results on this task \citep{ma2024multimodality}.


CPN achieves the best detection performance on CytoDArk0\_20x\_256, though it lags slightly behind CISCA in instance segmentation quality, as measured by bPQ, and across all metrics on CytoDArk0\_40x\_256. Both StarDist and CPN use cell representations as closed contours. However, while StarDist derives contour proposals from distance maps, CPN regresses contour parameters by anchoring full contours representations to the pixels within a cell. This, combined with its refinement mechanism, effectively alleviates StarDist's shape regularization problem.

CISCA and the CytoDArk0 dataset have been specifically developed to process histological WSIs of different species for comparative neuroanatomy studies, potentially aiding in the understanding of neurodegenerative and neuroinflammatory disorders. In this study, we evaluated CISCA on CytoDArk0 at 20x and 40x magnification, as well as on  the CoNIC challenge dataset of H\&E images of the colon,  the PanNuke pan-cancer dataset and the MoNuSeg dataset. Our comprehensive evaluations highlight CISCA’s robustness and accuracy in segmenting and classifying cells across diverse tissue types, magnifications, and staining methods. Future work should aim to expand the CytoDArk0 dataset to cover a wider variety of brain areas and species, potentially including cell classes that distinguish between neurons and glia. Additionally, exploring knowledge distillation from large foundation models into smaller, more efficient models like CISCA could enhance performance while avoiding overfitting and maintaining moderate computational requirements. These efforts would support a broad range of applications in digital pathology and comparative cytoarchitecture analyses.  Instance segmentation methods are tipically leveraged to extract morphological and topological features from individual cells, enabling the characterization of the tissue environment and supporting comparative analyses, diagnosis, prognosis or therapy response predictions tasks (cf. Section \ref{sec:Intro}). Additionally, they can be used to build \textit{cell-graphs}, where similar cells or cells in close proximity are linked, allowing to handle larger images than patch-based approaches and retain most of the contextual information \citep{meng2023clinical}. Cell-graphs can then be processed via \textit{graph neural networks} for improved disease classification or clustering \citep{zhou2019cgc,ramanathan2023self,vadori2024revealing}.

\section*{CRediT authorship contribution statement}
\textbf{Valentina Vadori}: Conceptualization, Methodology, Software, Formal Analysis, Investigation, Data Curation, Writing - Original Draft, Writing - Review \& Editing, Visualization.
\textbf{Jean-Marie Graïc}: Data Curation, Writing - Review \& Editing, Validation.
\textbf{Antonella Peruffo}: Data Curation, Writing - Review \& Editing, Validation.
\textbf{Giulia Vadori}: Data Curation, Writing - Review \& Editing.
\textbf{Livio Finos}: Writing - Review \& Editing, Validation.
\textbf{Enrico Grisan}: Resources, Writing - Review \& Editing,
Supervision, Project administration, Funding acquisition.



\normalsize
\bibliography{references}


\end{document}